\documentclass[journal]{IEEEtran}
\usepackage{cite}
\usepackage{times}
\usepackage{epsfig}
\usepackage{graphicx}
\usepackage{amsmath,bm}
\usepackage{amssymb}
\usepackage{epstopdf}
\usepackage{subfigure}
\usepackage{url}
\usepackage{multirow}
\usepackage{booktabs}
\usepackage{tabularx}

\begin{document}

\title{Learned Image Compression with Gaussian-Laplacian-Logistic Mixture Model
 and Concatenated Residual Modules}
%
%
%
\author{Haisheng~Fu,
        Feng~Liang,
        Jianping ~Lin,
        Bing ~Li,
        Mohammad ~Akbari,
        Jie~Liang,
        Guohe~Zhang,
        Dong Liu,
        Chengjie~Tu,
        Jingning Han
\thanks{Haisheng~Fu, Feng~Liang, Bing ~Li and Guohe~Zhang are with the School of Microelectronics, Xi'an Jiaotong University, Xi'an, China (e-mail: fhs4118005070@stu.xjtu.edu.cn; fengliang@xjtu.edu.cn; libing888@stu.xjtu.edu.cn; zhangguohe@xjtu.edu.cn) (Corresponding author: Feng Liang)}
\thanks{Jianping ~Lin  and Dong Liu are with the CAS Key Laboratory of Technology in Geo-Spatial Information Processing and Application System, University of Science and Technology of China, Hefei 230027, China (e-mail:  ljp105@mail.ustc.edu.cn; dongeliu@ustc.edu.cn)}
\thanks{ Mohammad Akbari and Jie Liang are with the School of Engineering Science, Simon Fraser University, Canada (akbari@sfu.ca; jiel@sfu.ca)}
\thanks{Chengjie~Tu is with the Tencent Technologies (e-mail: chengjietu@tencent.com)}
\thanks{Jingning~Han is with the Google Inc. (e-mail: jingning@google.com)}
\thanks{This work was supported by the National Natural Science Foundation of China (No. 61474093), the Natural Science Foundation of Shaanxi Province, China (No. 2021GXLH-Z-081), the Fundamental Research Funds for the Central Universities (No. xzd022020017), the Natural Sciences and Engineering Research Council of Canada (RGPIN-2020-04525), China Scholarship Council, and Google Chrome University Research Program.} }

\markboth{IEEE Transactions on Image Processing}%
{Shell \MakeLowercase{\textit{et al.}}: Bare Demo of IEEEtran.cls for IEEE Journals}
%

\maketitle

\begin{abstract}
Recently deep learning-based image compression methods have achieved significant achievements and gradually outperformed traditional approaches including the latest standard Versatile Video Coding (VVC) in both PSNR and MS-SSIM metrics. Two key components of learned image compression are the entropy model of the latent representations and the encoding/decoding network architectures. Various models have been proposed, such as autoregressive, softmax, logistic mixture, Gaussian mixture, and Laplacian. Existing schemes only use one of these models. However, due to the vast diversity of images, it is not optimal to use one model for all images, even different regions within one image. In this paper, we propose a more flexible discretized Gaussian-Laplacian-Logistic mixture model (GLLMM) for the latent representations, which can adapt to different contents in different images and different regions of one image more accurately and efficiently, given the same complexity. Besides, in the encoding/decoding network design part, we propose a concatenated residual blocks (CRB), where multiple residual blocks are serially connected with additional shortcut connections. The CRB can improve the learning ability of the network, which can further improve the compression performance. Experimental results using the Kodak,  Tecnick-100 and Tecnick-40 datasets show that the proposed scheme outperforms all the leading learning-based methods and existing compression standards including VVC intra coding (4:4:4 and 4:2:0) in terms of the PSNR and MS-SSIM. The source code is available at \url{https://github.com/fengyurenpingsheng}.

\end{abstract}


\IEEEpeerreviewmaketitle

\section{Introduction}

\begin{figure}[tb]
\centering
    \includegraphics[scale=0.3]{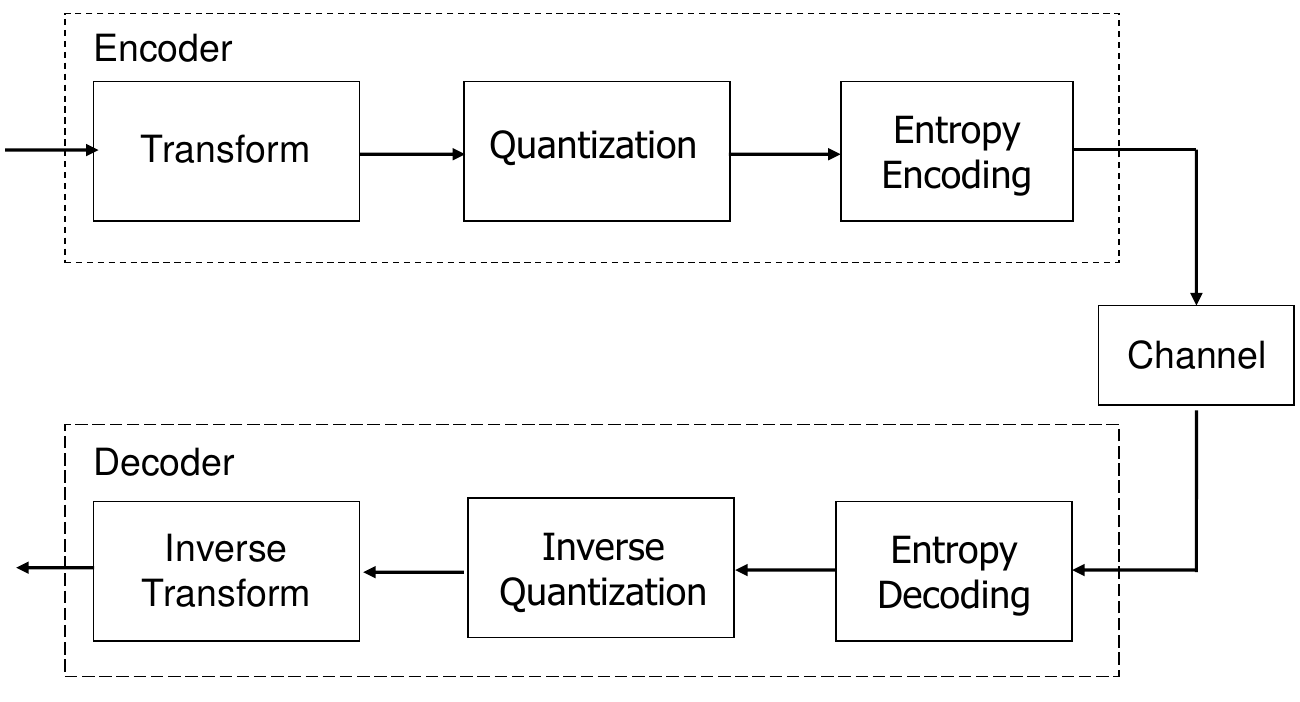}
	\caption{The block diagram of a typical image compression system.}
	\label{common_paradigm}
\end{figure}

\IEEEPARstart{I}{mage} compression is an important step in many applications. The classical approaches, e.g., JPEG \cite{JPEG}, JPEG 2000 \cite{JPEG2000}, and BPG (intra-frame coding of H.265/HEVC) \cite{BPG}, mainly use techniques such as linear transform, quantization, and entropy coding to remove the redundancy of the input and achieve better rate-distortion (R-D) performance, as illustrated in Fig. \ref{common_paradigm}. Recently, deep learning-based methods have been investigated, where the three main components are re-designed based on the properties of neural networks. This approach has gradually outperformed traditional methods in both PSNR and MS-SSIM metrics \cite{MS-SSIM}, and shows great potentials.

The most important difference of learning-based schemes is that the classic linear transform is replaced by a non-linear neural network, which is learned from training data. Therefore, how to design the network architecture to reduce the correlation of the latent representations, which are the output of the encoding network (also known as the feature maps), is critical to the performance of learned image compression schemes. Another important task is to design a good probability model to capture the remaining correlation of the latent representations, so that they can be encoded efficiently. 

\subsection{Encoder/Decoder Architectures}

\begin{table*}[!thp]
\caption{Different encoder/decoder architectures in image compression.}
\begin{center}
  \begin{tabular}{l|l}
  \hline
  \textbf{Methods} & \textbf{Highlights}  \\
  \hline
  RNN \cite{Toderici15,FullResolution} & The first work on end-to-end LSTM-based RNN for variable-rate image compression.\\
  
  GDN \cite{GDN} &GDN is first used in image compression framework, which  shows great potentials.\\
  
  Residual Network \cite{Lossy_image} &The residual network is first used for CNN-based image compression.\\
  
  GAN \cite{Rippel_2017} & The first to use GAN for image compression. A multi-scale framework is also proposed.\\
  
  Importance Map \cite{Limu_conf} & Importance Map is introduced to achieve content-aware bit allocation.\\
  
  Non-Local Attention Module \cite{chen2021} & A non-local attention module is introduced to capture long-range correlation.\\

  Simplified Attention Model \cite{cheng2020} & The non-local module in \cite{chen2021} is simplified. The Gaussian Mixture Model (GMM) is also used. \\

  Octave Convolution \cite{Mahammand_AAAI} &The multi-resolution octave convolution is employed in image compression framework.\\
  
  iWave++ \cite{Ma_2022_PAMI} & A lifting-based network similar to wavelet is proposed that supports both lossy and lossless compression. \\
  
  Transformer \cite{DCC2022} &The vision transformer is combined with convolutional layers to boost the compression performance.\\
    
\hline
\end{tabular}
\end{center}
\label{table1}
\end{table*}

\begin{table*}[!thp]
\caption{Different entropy coding models in image compression.}
\begin{center}
  \begin{tabular}{l|l}
  \hline
  \textbf{Methods} & \textbf{Highlights}  \\
  \hline
  
PixelCNN \cite{pixelCNN} &  Masked convolutional networks are used to predict the distribution of each pixel value. \\

  PixelCNN++ \cite{pixelCNN++} & The discretized logistic mixture model is used to reduce the complexity of PixelCNN.\\

3D-CNN \cite{Conditional} &The 2D-CNN model in \cite{pixelCNN++} is generalized to 3D. The importance map in \cite{Limu_conf} is also used.\\

Hyperprior \cite{Variational} & The hyperprior is first introduced. Zero-mean Gaussian model with a scale parameter is used.\\
  
  GMM and Autoregressive models\cite{Joint} & The GMM and autoregressive model are developed for the hyperprior method.\\

  Context Model \cite{Context_Model} & A context-adaptive entropy model is proposed with two types of contexts.\\

  Laplace-smoothed histogram \cite{Lossy_image} &The Laplace-smoothed histogram is used for entropy encoding.\\

  Laplacian \cite{tuya_2019} &The Laplacian distribution is established for lossy compression.\\
  
  Logistic Mixture  \cite{L3C} & The discretized logistic mixture distribution is used.\\
  
GMM and quality enhancement network \cite{Lee_2021} & The GMM model and a quality enhancement subnetwork are used to achieve the state of the art.\\

Multivariate GMM \cite{Zhu_2022_CVPR} & Multivariate GMM and vector quantization are used. The parameters are estimated in a cascaded approach.\\

\hline
\end{tabular}
\end{center}
\label{table2}
\end{table*}

Table \ref{table1} summarizes some representative encoder/decoder architectures for learned image coding. 

One of the first learned image compression schemes was proposed in \cite{Toderici15}, which was based on the long short-term memory (LSTM) recurrent neural network (RNN) and was used to compress thumbnail images. In \cite{FullResolution}, the scheme in \cite{Toderici15} is generalized to full-resolution images.

In \cite{GDN, end_to_end}, an encoder network that includes convolution, downsampling, and the generalized divisive normalization (GDN) is proposed. It was the first learned image compression scheme that achieved better performance than JPEG2000 in terms of both PSNR and MS-SSIM.

Most recent schemes are based on the autoencoder framework, whose latent representations have much lower dimension than the input, and is very suitable for data compression. 

A powerful building block of many cutting-edge neural networks is the residual block first proposed in the ResNet \cite{resblock}, which uses shortcut connections to facilitate the design and training of deep networks, and can effectively improve the performances of many computer vision tasks. As a result, the residual block has also been used in many learning-based image compression schemes.

In \cite{Lossy_image}, the residual block concept in \cite{resblock} was used in the autoencoder architecture, which also achieved comparable performance to JPEG2000.

Generative adversarial network (GAN) is another powerful framework for many applications, and has also been used in several learned image coding schemes \cite{Rippel_2017, Santurkar_2018, DSSLIC}.

In \cite{Limu_conf}, the importance map is introduced to achieve content-adaptive bit allocation in different regions, but the importance mask needs to be sent to the decoder. In \cite{chen2021}, non-local attention module is used to capture long-range correlation, and generate attention mask without sending side information, and it is used in both core network and hyper network. In \cite{cheng2020}, the attention module in \cite{chen2021} is simplified.

In \cite{Lin_MMSP, Mahammand_AAAI}, a multi-resolution network architecture based on the octave convolution in \cite{octave} is developed, similar to the wavelet transform. In \cite{Ma_2022_PAMI}, another wavelet-like scheme is proposed based on the lifting scheme, which supports both lossy and lossless compression.
  
In \cite{DCC2022}, the vision transformer framework is introduced and combined with convolutional layers to boost the image compression performance.

\subsection{Entropy Coding Models}

Table \ref{table2} is an overview of typical entropy coding models used for learned image coding. 

In earlier schemes \cite{soft_to_hard, Lossy_image, end_to_end},  the quantized latent representations or latents are assumed to be independent and identically distributed and followed a simple marginal distribution. Once the parameters of the distribution are trained, the probability of all latents is fixed for all images, which is used by the entropy encoding and decoding of the quantized latents. Since a fixed entropy model is used, the performances of these methods are compromised.

To improve the entropy coding performance, recent methods consider the correlation of neighboring pixels or latents (contexts), and achieve image-adaptive entropy coding. To estimate the joint probability with manageable complexity, the chain rule is usually used, which factorizes the joint probability into products of conditional probabilities.

In PixelCNN \cite{pixelCNN}, masked convolutional networks are used to predict the distribution of each pixel value, conditioned on the causal neighbors in an autoregressive manner, which is used for image generation and image decoder in an autoencoder. In \cite{pixelCNN++}, the PixelCNN++ is proposed, which uses the discretized logistic mixture distribution to reduce the complexity of \cite{pixelCNN}. In \cite{L3C}, PixelCNN++ is applied to lossless image compression. 

To help the estimation of the conditional probabilities of the latents, the hyperprior network is introduced in \cite{Variational}, where a hyper encoder network is used to extract some hyperpriors from the latents, which are coded as side information and sent to a hyper decoder network. The latter uses the reconstructed hyperpriors to estimate the conditional probability of the latents, thereby making the entropy model image-dependent and spatially adaptive. 

In \cite{Variational}, the conditional probability of each latent is assumed to follow a zero-mean Gaussian scale mixture (GSM) model, which achieves better performance than BPG (4:4:4). In \cite{Joint}, the non-zero-mean Gaussian mixture model (GMM) is further proposed. In addition, the autoregressive context model in \cite{pixelCNN} is introduced to estimate the conditional probability of each latent from both the hyperpriors and its spatial context.

Other models have also been proposed to encode the latents. For example, in \cite{tuya_2019}, the Laplacian distribution was used. In \cite{cheng2020} and \cite{Lee_2021}, the GMM model is adopted. A quality enhancement subnetwork is also introduced in \cite{Lee_2021} to improve the performance. It is the first learned image compression approach that achieves better performance than H.266/VVC (4:4:4) in both PSNR and MS-SSIM. As far as we know, the results in \cite{Lee_2021} are currently the best in the literature.

\subsection{Multivariate Models}

The entropy coding approach above predicts the conditional probability of a latent given some causal neighbors and the hyperpriors, according to the chain rule, which approximates the joint probability implicitly. Another possible approach is to use the more general multivariate models to represent the joint probability explicitly.

In \cite{Zhu_2022_CVPR}, the multivariate GMM is used in learned image compression to capture the inter-channel correlation of co-located latents across feature maps (this is captured implicitly by the approach above using 1x1 filters in the hyper decoder). The means and covariance matrices are estimated in a cascaded approach. Vector quantization is used to encode the latents, as in \cite{soft_to_hard}. The hyper coding network is not used. However, since no spatial context is used, the R-D performance of \cite{Zhu_2022_CVPR} is only comparable to VVC on the Kodak test set, and not as good as other leading methods such as \cite{Lee_2021}. This indicates that if the learned image compression framework can consider the joint distribution via context model, non-local operator, and attention module, it is sufficient to use univariate models to model the remaining redundancy in the latents.

On the other hand, the overall complexity of multivariate model-based methods is usually quite high, as noted in \cite{Zhu_2022_CVPR}, because the number of parameters of multivariate models increases quadratically with the size of the covariance matrix. Therefore, the complexity of the network model and the complexity of calculating multivariate probability increase dramatically. The training of the network is also more difficult, because the reverse gradients of different variables affect each other. Although the method in \cite{Zhu_2022_CVPR} can be parallelized because no spatial context is used, it still has very high demands for GPU and memory. In fact, other methods using spatial context can also be parallelized \cite{He_2021_CVPR,Li_2020_TIP,lu2022high}.

Multivariate models have also been used in other applications. Most of them are based on the Gaussian distribution. In \cite{Viroli}, a deep GMM is proposed for classification by using the factor-analytic representation of the GMM in each layer. The Expectation-Maximization (EM) algorithm is used to estimate the parameters. However, it is found in \cite{BumpyGaussian} that sometimes it is quite challenging for this method to infer its parameters, even for a small-scale problem with 3 layers and 76000 latents.

In \cite{Hao_multivariate_2021},  a novel face recognition framework based on the multivariate GMM is proposed to make feature embeddings extract more identity-relevant information. In \cite{Xu_multivariate_2021}, a learned multiple graph Gaussian embedding model is developed to learn highly informative network features by mapping high-dimensional networks into a low-dimensional latent space. In \cite{Koray_multivariate_2022}, the multivariate skewed t-distribution is proposed for hyperspectral anomaly detection. In \cite{GMM_ImageRecognition}, the last fully-connected (FC) layer of a deep network is combined with a mixture of GMM (MoGMM) for image recognition. In \cite{GMM_ImageRestoration}, the multivariate GMM is considered as a prior to recover degraded images. In \cite{Arslan_ECCV_Gaussian}, a framework for deep unconstrained face verification is proposed to map learned discriminative facial features to a regularized metric space, in which matching and non-matching pairs follow multivariate Gaussian distributions.

\subsection{Contributions of this Paper}

Although the learned methods have made significant progresses, the compression performance can be further improved. First, the previous entropy models only use a single distribution, which is not optimal for every latent. Second, the latents still exist some spatial redundancy. In order to address these issues, we make the following contributions in this paper:

$\bullet\,$ Instead of using a single probability model, we propose a discretized Gaussian-Laplacian-Logistic mixture model (GLLMM), which is more flexible and efficient in estimating the conditional probabilities of the latents, given the same complexity. In fact, the previous models are special cases of the proposed model. Ablation studies demonstrate that the proposed joint model outperforms all previously proposed single models. For example, it achieves 0.1-0.15 dB gain in PSNR compared to the GMM model for the Kodak dataset.

$\bullet\,$ We improve the basic building block of the encoder network by developing a concatenated residual module (CRM), with additional shortcut connections. The CRM improves the information flow, reduces the correlation of the output, and improves the training of the network. Ablation studies using the Kodak dataset show that the proposed CRM can achieve 0.2-0.3 dB gain in terms of PSNR compared to the original residual blocks. We also conduct ablation studies to compared the performances of CRMS with two and three stages of residual blocks.

$\bullet\,$ One of the key contributions of \cite{Lee_2021} is the post-processing component, which achieves 0.5 dB gain in \cite{Lee_2021} compared to its baseline. We also investigate the performance of this component in our scheme. However, our experiments show that there is no need to apply the post-processing in our method, because the proposed CRM and entropy model have done a good job.  This also reduces the complexity of the scheme.

Experimental results using the Kodak\cite{Kodak}, Tecnick-100, and Tecnick-40 datasets in \cite{Tecnick} show that the proposed scheme outperforms all the previous learning-based methods including  \cite{Lee_2021,cheng2020} and traditional codecs including VVC in both PSNR and MS-SSIM. For example, in terms of PSNR, for the Kodak dataset,  when the bit rate is higher than 0.4 bpp, our method is 0.2-0.3 dB higher than \cite{Lee_2021, cheng2020}, and 0.3-0.4 dB higher than VVC (4:4:4). For the Tecnick dataset, when the bit rate is higher than 0.2 bpp, our method is 0.3-0.4 dB better than VVC (4:4:4). This represents the new state of the art in learned image compression.


The remainder of the paper is organized as follows.  In Section  \ref{The Proposed structure}, we propose our image compression framework with discretized Gaussian-Laplacian-Logistic mixture model and the concatenated residual modules. In Sec. \ref{Experiment}, we compare our method with some state-of-the-art learning-based methods and classical image compression methods. Ablation experiments are carried out to investigate the performance gain of the the proposed scheme. The conclusions are reported in Sec. \ref{Conclusion}.

\begin{figure*}[tb]
	\centering
		\includegraphics[scale=0.35]{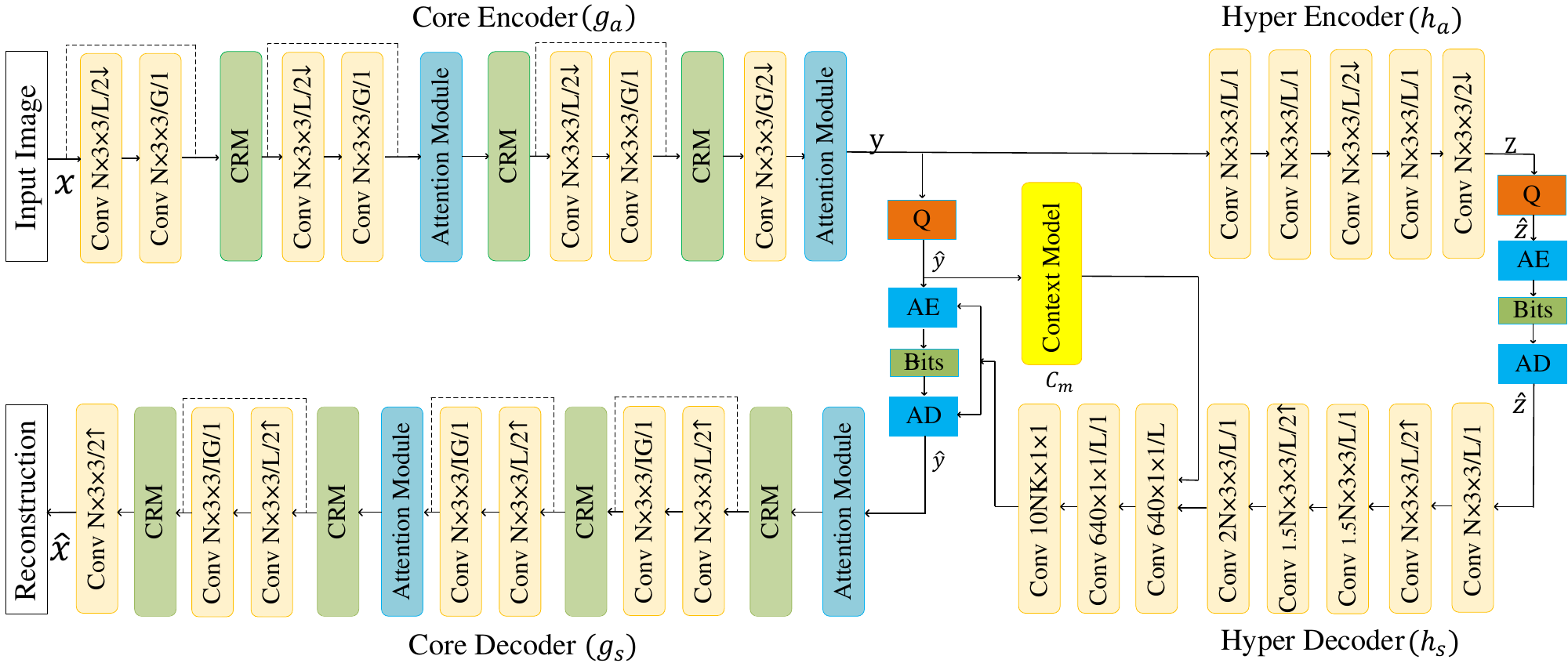}
	\caption{The Framework of the proposed image compression scheme. $G$ and $IG$ represent the GDN module and inverse GDN module, respectively. $\uparrow$ and $\downarrow$ denote the up- or down- sampling.  $3 \times 3$ is the size of convolution kernel. $Q$ represents quantization. $AE$ and $AD$ represent the arithmetic encoder and arithmetic decoder, respectively. $L$ represents leaky ReLU. The dotted lines denote the shortcut connection with size change, as in \cite{resblock,cheng2020}.}
	\label{networkstructure}
\end{figure*}

\section{The Proposed Image Compression Scheme with CRM and GLLMM}
\label{The Proposed structure}

In this section, we first introduce the overall learned image coding framework adopted in this paper, and then describe the proposed concatenated residual module (CRM) and the discretized Gaussian-Laplacian-Logistic mixture model (GLLMM).

\subsection{The Overall Framework}

The learned image coding framework adopted in this paper is illustrated in Fig. \ref{networkstructure}, which is mainly based on \cite{cheng2020}. It is comprised of two main parts: the core autoencoder and the hyperprior coding.

The size of the input color image $\bm{x}$ is $W\times H\times 3$,  where $W$ and $H$ represents the width and height of the image, respectively. In the training of the network, the pixel values are normalized to $[-1, 1]$ by ($\frac{\bm{x}}{127.5} - 1.0$).

The core autoencoder includes an encoder ($g_{a}$) and a decoder ($g_{s}$). The input image $\bm{x}$ is first sent to the encoder network $g_{a}$, which aims to reduce the redundancy and learn a compact latent representation $\bm{y}$ of the input image. The latent is then quantized and entropy coded. The quantized latent is denoted by $\hat{\bm{y}}$.

The core encoder network includes various convolution layers and four stages of pooling operators to get the latents. The residual blocks with shortcut connections are used extensively. The GDN operator is used when the size is changed.

To improve the R-D performance, the simplified attention module in \cite{cheng2020} is applied at two resolutions to capture long-range correlation, and strengthen more important areas, so that they will get more bit allocation.

To estimate the distribution of the latents and improve entropy coding efficiency, the hyperprior coding part is introduced in \cite{Variational}, which consists of a hyper encoder ($h_{a}$) and a hyper decoder ($h_{s}$). The hyper encoder extracts the hyperprior $\bm{z}$ from the latents, which is quantized into $\hat{\bm{z}}$ and entropy coded as a side information to the hyper decoder. The hyper decoder first recovers $\hat{\bm{z}}$ via entropy decoding, and then uses hyper decoder network $h_{s}$ to estimate the parameters of the \text{conditional} distribution of $\hat{\bm{y}}$, which is used by the entropy encoding and decoding of $\hat{\bm{y}}$.

Since the input dimension of the hyper network is much smaller than the original image, the hyper network is much simpler than the core network. Leaky ReLU is utilized in most convolution layers, except for the last layer in hyper encoder and decoder, which do not have any activation function.

To further improve the entropy coding efficiency, the context model network $c_{m}$ in \cite{Joint} and \cite{cheng2020} is also used in our system, which uses masked convolutions to capture the correlation of neighboring latents. The output layer of the context model is concatenated with the output of the first part of hyper decoder, and then further processed by some additional convolutional layers to estimate the parameters of the conditional distributions of the latents, which are then used to entropy encode and decode $\hat{\bm{y}}$.

Since autoregressive context model is used, the symbols have to be decoded in serial manner. After all the symbols of $\hat{\bm{y}}$ are decoded, they are sent to the core decoder ($g_{s}$) to generate the reconstructed image. Note that the context model can be sped up using the methods in  \cite{He_2021_CVPR,Li_2020_TIP,lu2022high}.


In Fig. \ref{networkstructure}, the encoders and decoders in the core and hyper networks are symmetric, except that they use convolution and deconvolution operators, respectively.




\subsection{The Proposed Concatenated Residual Module (CRM)}

\begin{figure}[tb]
\centering
\subfigure[]{
\begin{minipage}[t]{0.5\linewidth}
\centering
\includegraphics[scale=0.5]{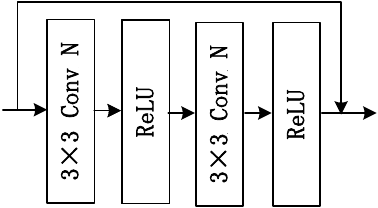}
\label{residual_a}
\end{minipage}
}%
\subfigure[]{
\begin{minipage}[t]{0.5\linewidth}
\centering
\includegraphics[scale=0.5]{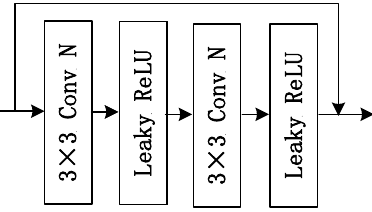}
\label{residual_b}
\end{minipage}
}%

\subfigure[]{
\begin{minipage}[t]{1\linewidth}
\centering
\includegraphics[scale=0.6]{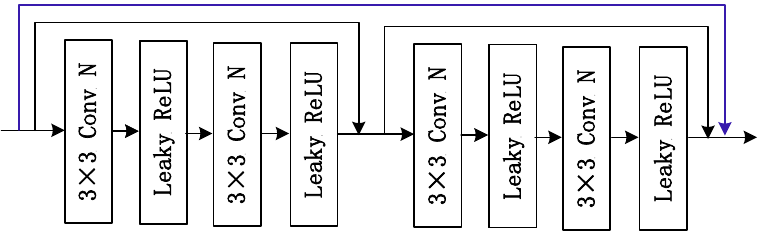}
\label{residual_c}
\end{minipage}
}%

\subfigure[]{
\begin{minipage}[t]{1\linewidth}
\centering
\includegraphics[scale=0.3]{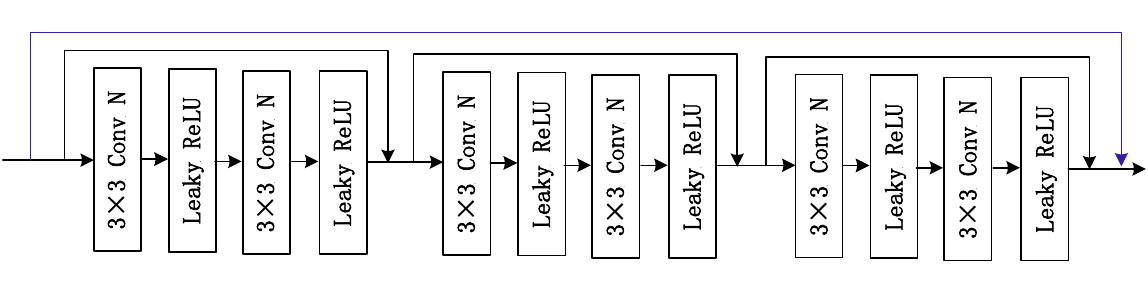}
\label{residual_d}
\end{minipage}
}%
\caption{(a) The standard residual block; (b) the residual block with Leaky ReLU; (c) the proposed two-stage concatenated residual module; (d) the proposed three-stage concatenated residual module.}
\label{DRBCM}
\end{figure}

When the size does not change, to further remove the spatial correlation in the latent representation, we develop two deeper residual blocks in this paper, which is illustrated in Fig. \ref{DRBCM}. The basic building block is the standard residual block developed in the ResNet \cite{resblock}, as described in Fig. \ref{residual_a}. As in \cite{cheng2020}, we first employ the leaky ReLU activation function to replace the ReLU function and remove the batch normalization layer from the residual block. The leaky ReLU activation function can speed up the convergence of the network. The detailed structure is shown in Fig. \ref{residual_b}, which is used in \cite{cheng2020}. Based on this, we develop two concatenated residual modules, as shown in Fig. \ref{residual_c} and Fig. \ref{residual_d}.

In Fig. \ref{residual_c}, two residual blocks in Fig. \ref{residual_b} are concatenated, and another short connection is added between the input and the output. In Fig. \ref{residual_d}, three residual blocks are concatenated, with an additional shortcut connection as well.

Compared to the standard residual block in Fig. \ref{residual_b}, the concatenated modules have larger receptive fields. They can remove more spatial correlation, which can also help the attention module in the network. In Sec. III, we will compare the performances of the two types of concatenated modules.

\subsection{The Proposed Gaussian-Laplacian-Logistic Mixture Model (GLLMM)}

\subsubsection{Existing Models}

The core encoder and decoder in Fig. \ref{networkstructure} can be represented by
\begin{equation}\label{single_image_compression_Eq}
\begin{split}
   \bm{y} &= g_{a}(\bm{x;\phi}), \\
   \hat{\bm{y}}&= Q(\bm{y}),\\
   \hat{\bm{x}}&= g_{s}(\hat{\bm{y}}; \bm{\theta}),
\end{split}
\end{equation}
where $\bm{\phi}$ and $\bm{\theta}$ are the parameters of the encoder and decoder networks that need to be optimized. $Q$ represents the quantization operator. 

In earlier work, a fixed distribution is used to encode all entries of $\hat{\bm{y}}$ for all images, which is not optimal. In \cite{Variational}, the hyper encoder $h_a$ and hyper decoder $h_s$ are introduced to help learning the distribution of each entry of $\hat{\bm{y}}$, which makes the entropy coding image-adaptive. The output $\bm{z}$ of the hyper encoder network is quantized into $\hat{\bm{z}}$ and entropy coded as a side information to the decoder. The hyper coding part can be represented by
\begin{equation}\label{Variational_Eq}
\begin{split}
   \bm{z} &= h_{a}(\bm{y};\bm{\phi}_{h}), \\
   \hat{\bm{z}}&= Q(\bm{z}),\\
   P_{\hat{\bm{y}}| \hat{\bm{z}}}(\hat{\bm{y}}|\hat{\bm{z}})& \leftarrow h_{s}(\hat{\bm{z}}; \bm{\theta}_{h}).\\
\end{split}
\end{equation}
where $\bm{\phi}_{h}$ and $\bm{\theta}_{h}$ are the parameters of the hyper encoder and hyper decoder, and $P_{\hat{\bm{y}}|\hat{\bm{z}}}(\hat{\bm{y}}|\hat{\bm{z}})$ is the conditional distribution vector of $\hat{\bm{y}}$ given $\hat{\bm{z}}$.

In \cite{Variational}, $P_{\hat{\bm{y}}|\hat{\bm{z}}}(\hat{\bm{y}}|\hat{\bm{z}})$ is assumed to follow the independent zero-mean Gaussian distribution with variation vector $\bm{\sigma}^2$. In \cite{Joint}, it is allowed to have non-zero-mean Gaussian distribution with mean vector $\bm{\mu}$:
\begin{equation}\label{GLL_Eq}
   P_{\hat{\bm{y}}|\hat{\bm{z}}}(\hat{\bm{y}}|\hat{\bm{z}})  \sim \mathcal{N} (\bm{\mu}, \bm{\sigma}^{2}).
\end{equation}

In addition, the autoregressive context model $C_{m}$ in\cite{pixelCNN} is introduced in \cite{Joint} to consider the correlation from causal neighboring latents in estimating the distribution of the latent.

Based on \cite{Joint}, in \cite{cheng2020}, a Gaussian mixture model (GMM) is further introduced to estimate the latents. Since context model is also used as in \cite{Joint}, the conditional probability is denoted by  $P_{\hat{\bm{y}} | \hat{\bm{y}}_c, \hat{\bm{z}}}(\hat{\bm{y}}|\hat{\bm{y}}_c,\hat{\bm{z}})$, where $\hat{\bm{y}}_c$ represents the neighboring latents (contexts) for $\hat{\bm{y}}$. Therefore the GMM-based estimation can be written as
\begin{equation}\label{GMM_Eq}
   P_{\hat{\bm{y}} | \hat{\bm{y}}_c, \hat{\bm{z}}}(\hat{\bm{y}} | \hat{\bm{y}}_c, \hat{\bm{z}})  
   \sim \sum_{k=1}^K{ \bm{w}^{(k)} \mathcal{N}  (\bm{\mu}^{(k)}, \bm{\sigma}^{2(k)})}.
\end{equation}

At each position $i$, the conditional distribution of $\hat{y}_i$, $P_{\hat{\bm{y}}|\hat{y}_{i,c}, \hat{\bm{z}}}(\hat{y}_i|\hat{y}_{i,c}, \hat{\bm{z}})$, is a weighted average of multiple Gaussian models with different means and variances (the sum of the weights equals to $1$). This is more powerful than a single Gaussian model.

\begin{figure}[tb]
\centering
\includegraphics[scale=0.5]{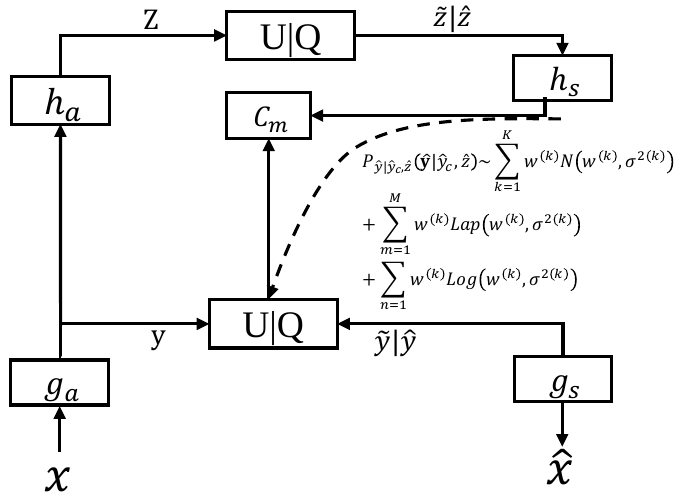}
\centering
\caption{Illustration of the proposed entropy coding model.}
\label{gllmm_model}
\end{figure}


Models with other distributions have also been investigated, such as logistic mixture model in \cite{pixelCNN++, L3C}, and Laplacian distribution in \cite{tuya_2019}.

However, the aforementioned methods only use one type of probability models to learn the probability distribution of the latent representation. It is difficult for one probability model to learn the distributions of all images. Therefore, better performance can be expected if different types of distributions can be combined together.

\begin{figure*}[t]
\centering
\subfigure
{\begin{minipage}[t]{0.5\linewidth}
\centering
\includegraphics[width=\columnwidth]{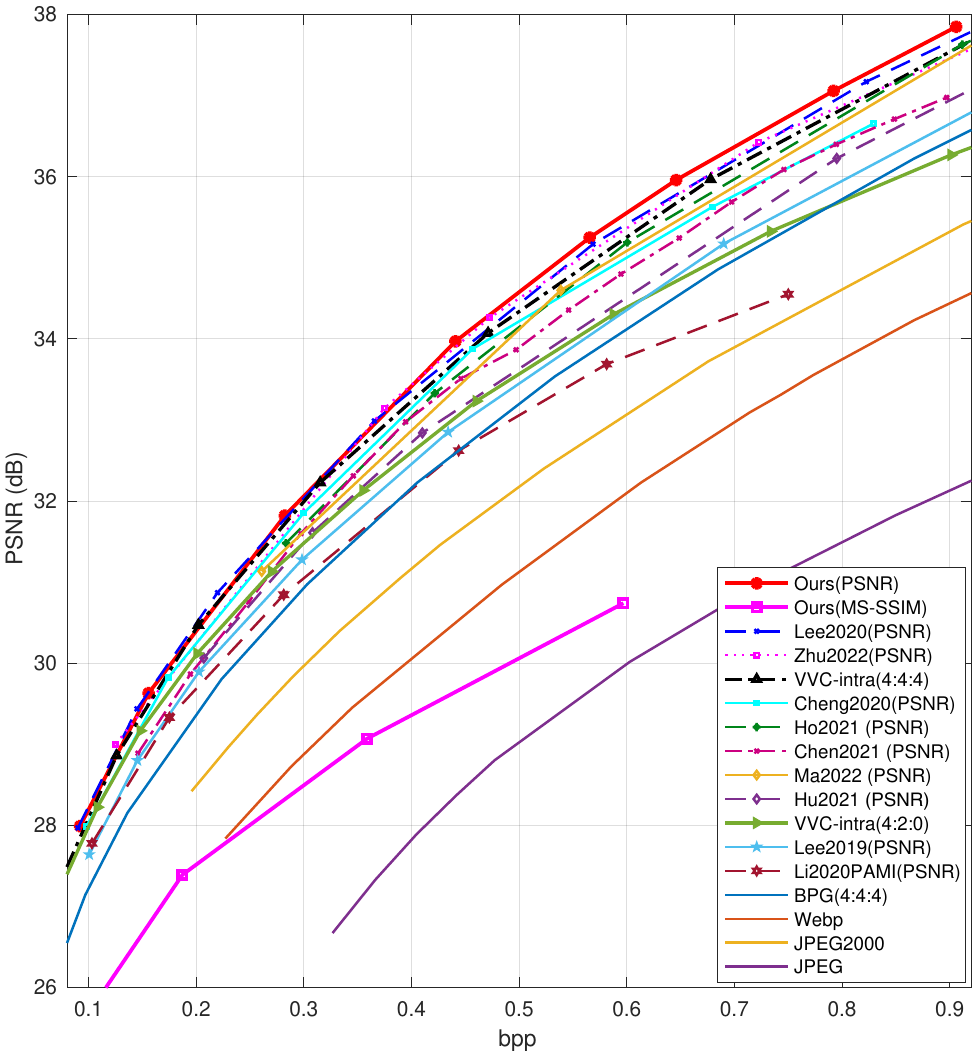}
\end{minipage}
}%
\subfigure
{\begin{minipage}[t]{0.5\linewidth}
\centering
\includegraphics[width=\columnwidth]{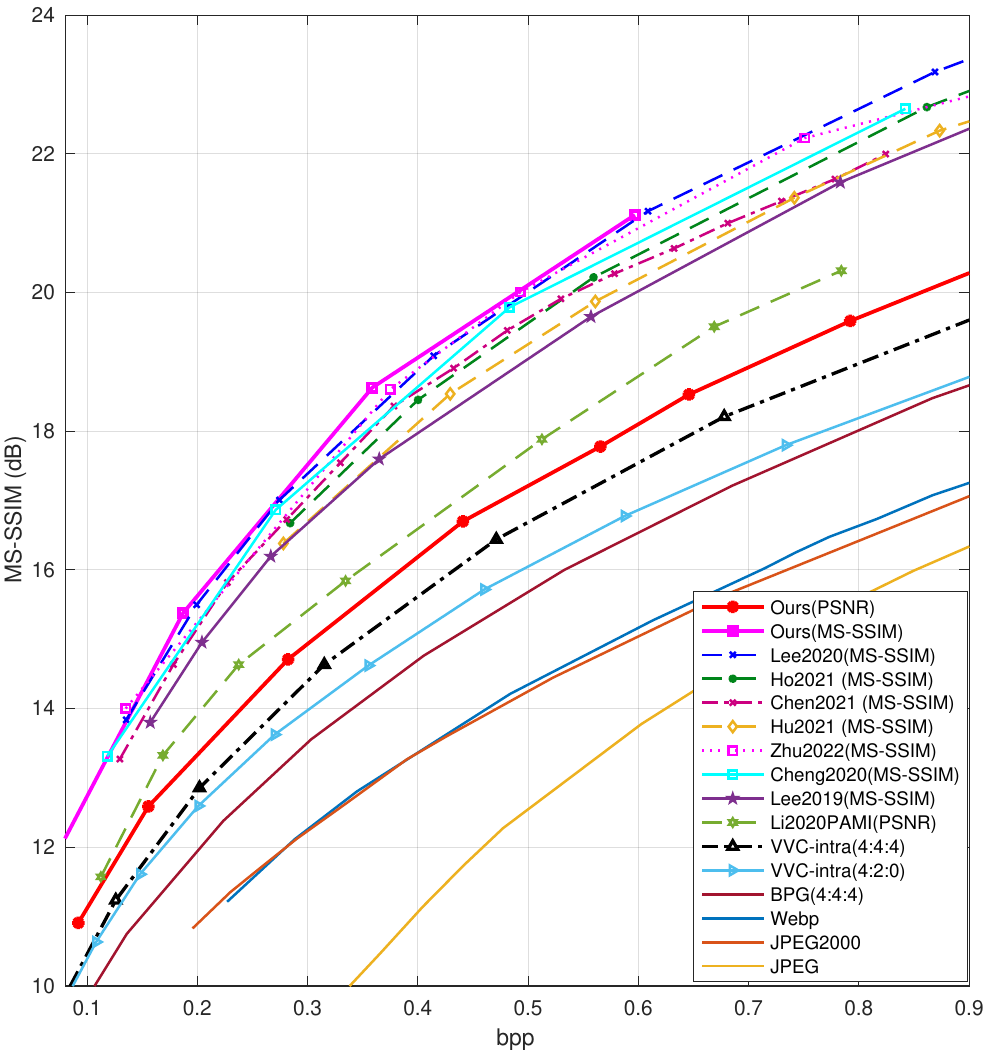}
\label{kodak_SSIM}
\end{minipage}
}%
\centering
\caption{The average PSNR and MS-SSIM performances of different methods on all 24 Kodak images.}
\label{test_kodak}
\end{figure*}

\begin{figure*}[t]
\centering
\subfigure
{\begin{minipage}[t]{0.5\linewidth}
\centering
\includegraphics[width=\columnwidth]{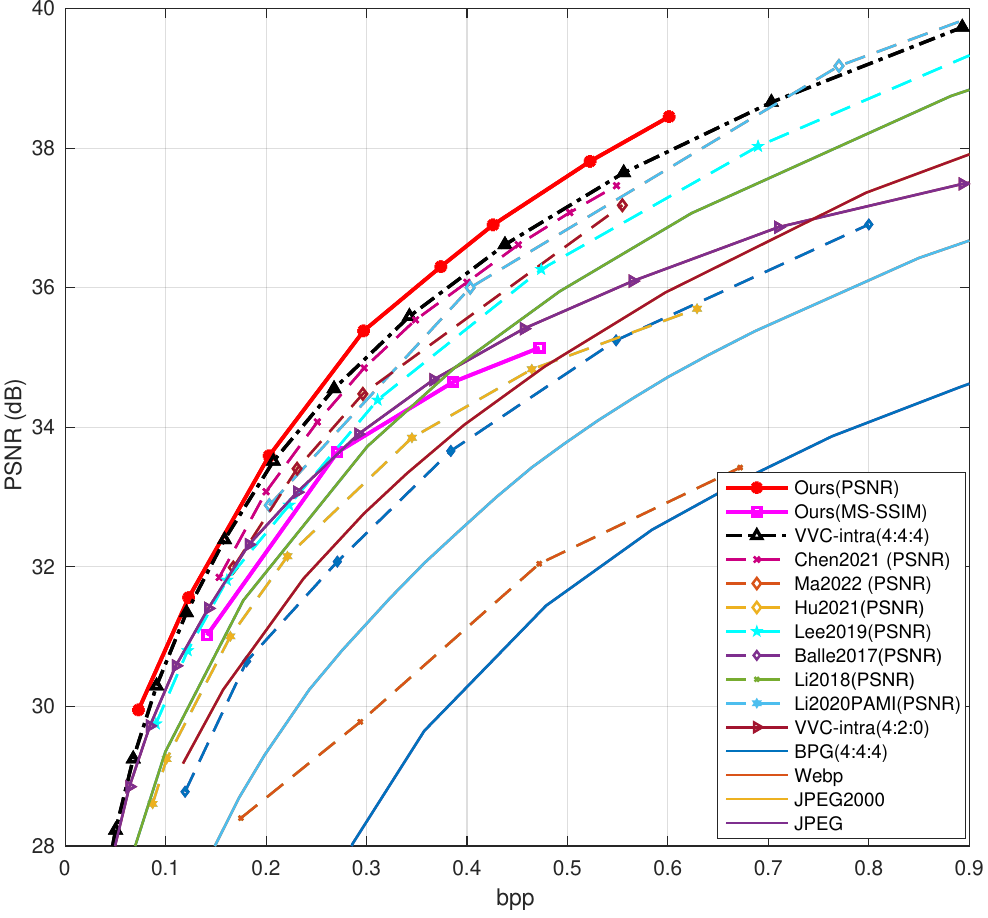}
\end{minipage}
}%
\subfigure
{\begin{minipage}[t]{0.5\linewidth}
\centering
\includegraphics[width=\columnwidth]{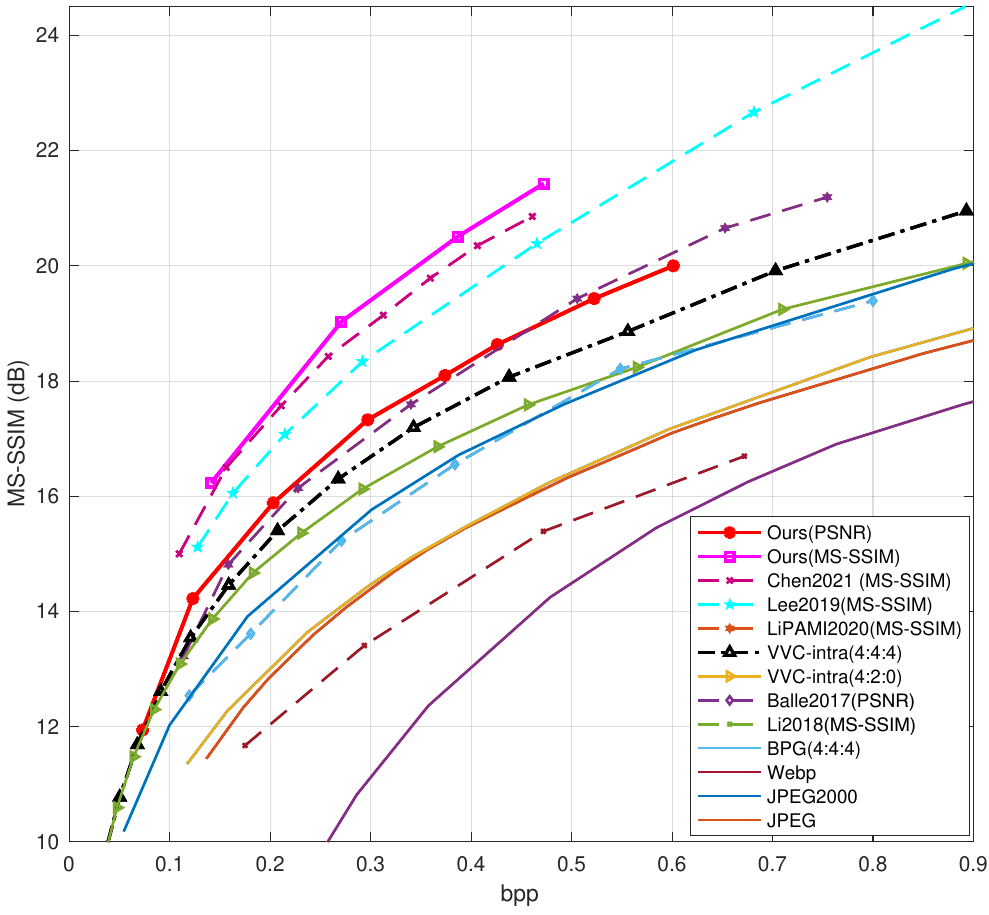}
\end{minipage}
}%
\centering
\caption{The average PSNR and MS-SSIM performances of different methods on all Tecnick-100 images.}
\label{test_Tecnick}
\end{figure*}

\begin{figure*}[t]
\centering
\subfigure
{\begin{minipage}[t]{0.5\linewidth}
\centering
\includegraphics[width=\columnwidth]{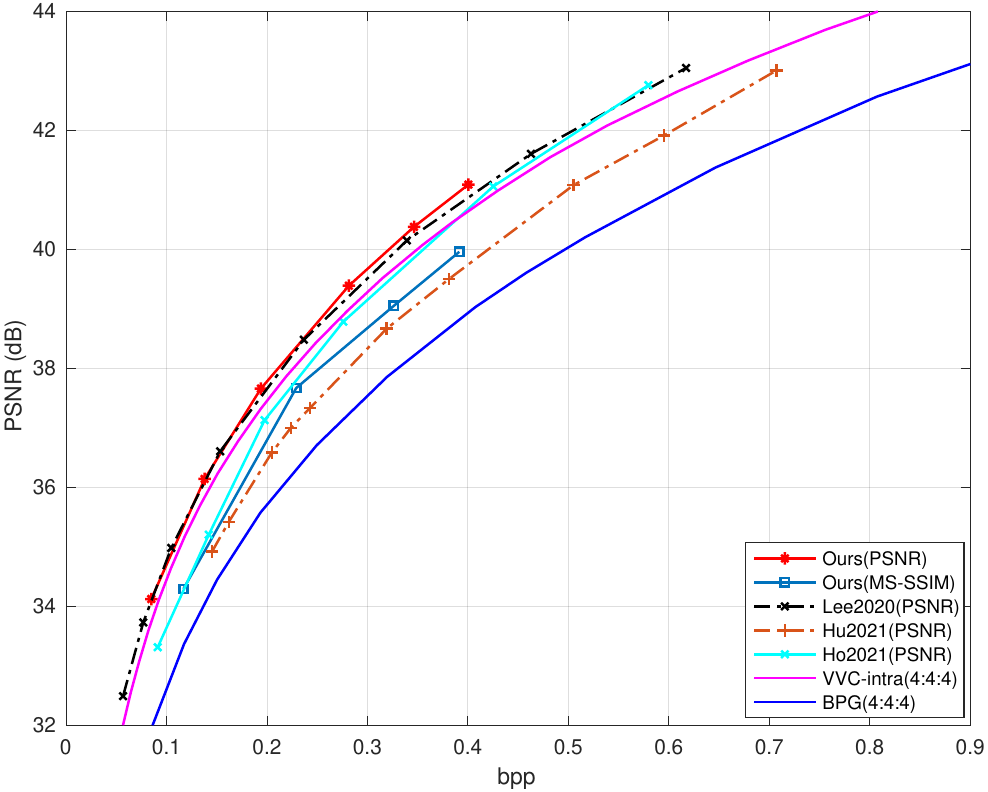}
\end{minipage}
}%
\subfigure
{\begin{minipage}[t]{0.5\linewidth}
\centering
\includegraphics[width=\columnwidth]{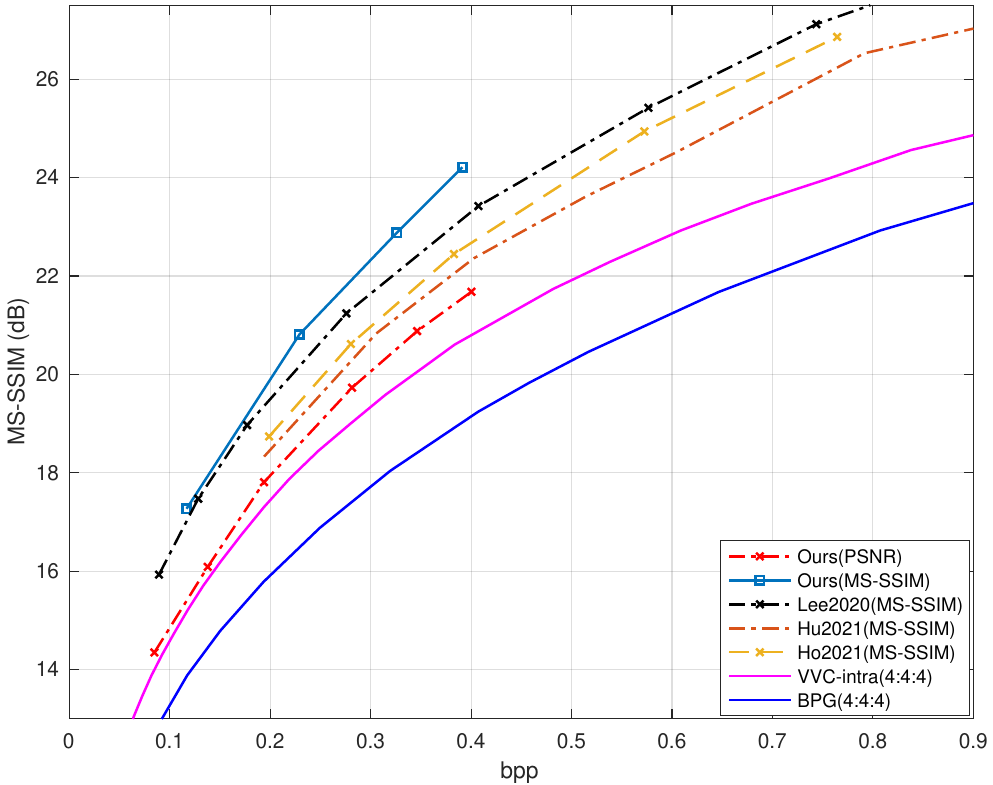}
\end{minipage}
}%
\centering
\caption{The average PSNR and MS-SSIM performances of different methods on all Tecnick-40 images.}
\label{test_Tecnick40}
\end{figure*}

\subsubsection{The Proposed GLLMM Model}

In this paper, we propose a powerful Gaussian-Laplacian-Logistic model (GLLMM) distribution as shown in Fig. \ref{gllmm_model} for $P_{\hat{\bm{y}}|\hat{\bm{y}}_c,\hat{\bm{z}}}(\hat{\bm{y}}|\hat{\bm{y}}_c,\hat{\bm{z}})$:
\begin{equation}\label{GLLMM_Eq}
\begin{split}
   P_{\hat{\bm{y}}|\hat{\bm{y}}_c,\hat{\bm{z}}}(\hat{\bm{y}}|\hat{\bm{y}}_c,\hat{\bm{z}}) & \sim \bm{p}_{0} \sum_{k=1}^K{\bm{w}^{(k)} \mathcal{N}  (\bm{\mu}_w^{(k)}, \bm{\sigma}_w^{2(k)})}\\
   &+ \bm{p}_{1} \sum_{m=1}^M{\bm{\alpha}^{(m)}Lap(\bm{\mu}_{\alpha}^{(m)}, \bm{\sigma}_{\alpha}^{2(m)})}\\
   &+ \bm{p}_{2} \sum_{j=1}^J{\bm{\beta}^{(j)}Logi(\bm{\mu}_{\beta}^{(j)}, \bm{\sigma}_{\beta}^{2(j)})}, \\
\end{split}
\end{equation}
which is a weighted average of Gaussian mixture, Laplacian mixture ($Lap$), and logistic mixture ($Logi$), with normalization constraints for $\bm{w}^{(k)}$, $\bm{\alpha}^{(m)}$, $\bm{\beta}^{(j)}$, and $\bm{p}_i$ respectively. $\bm{\mu}$ and $\bm{\sigma}$ with different subscripts represent the mean and variance parameters for different models, and $\bm{p}_{i}$'s are the weights of the three types of distributions. 

The total number of parameters in Eq. (\ref{GLLMM_Eq}) is $3(K+M+J+1)$. Each latent needs its own set of parameters to estimate its conditional distribution. All of these parameters are generated by the hyper decoder network. The last layer of the hyper decoder uses a $1\times1$ filter to estimate the distribution parameters of each latent. The kernel size of each $1\times1$ filter is thus $3(K+M+J+1)$. For $N$ feature maps, the total size of the $1\times1$ filter at each position is  $3N(K+M+J+1)$. In this paper, the values of $K$, $M$, and $J$ are all chosen to be $3$, based on the experimental results in Sec. III. Therefore there are $30$ parameters to estimate for each latent, and the total size of the $1 \times 1$ filter for all $N$ feature maps are $30N$, as shown in Fig. \ref{networkstructure}.

Compared to the previous models with a single distribution, the proposed GLLMM model includes three types of distributions, and can capture the distribution of the latent more accurately and efficiently, given the same complexity, thereby improving the performance.

Eq. (\ref{GLLMM_Eq}) is the continuous distribution, but the entropy coding part needs the distribution of the quantized $\hat{\bm{y}}$. Since quantization is not differentiable, a standard solution during training is to add a uniform noise $U(-\frac{1}{2},\frac{1}{2})$ to $\bm{y}$ to achieve a differentiable approximation of the quantization step, which enables the back-propagation-based training \cite{Variational}. During the inference, $\bm{y}$ is quantized to $\hat{\bm{y}}$ as usual, which is then encoded into the bit stream via entropy coding.

Based on the approach above, the distribution of the discrete-valued $\hat{\bm{y}}$ after quantization is given by \cite{cheng2020}
\begin{equation}\label{GLLMM_Eq_discretized}
\begin{split}
   P_{\hat{\bm{y}}|\hat{\bm{y}}_c,\hat{\bm{z}}} & (\hat{\bm{y}}|\hat{\bm{y}}_c,\hat{\bm{z}}) = \prod_{i}(P_{\hat{\bm{y}}|\hat{\bm{y}}_c,\hat{\bm{z}}}(\hat{y}_i|\hat{y}_{i,c},\hat{\bm{z}})),\\
   P_{\hat{\bm{y}}|\hat{\bm{y}}_c,\hat{\bm{z}}} & (\hat{y}_i|\hat{y}_{i,c},\hat{\bm{z}}) = \left[ \left(
   p_{0,i} \sum_{k=1}^K{w_{i}^{(k)} \mathcal{N}  \left(\mu_{w,i}^{(k)}, \sigma_{w,i}^{2(k)}\right)} \right. \right.\\
   &+p_{1,i} \sum_{m=1}^M{\alpha_{i}^{(m)}Lap \left(\mu_{\alpha,i}^{(m)}, \sigma_{\alpha,i}^{2(m)}\right)}\\
   & \left. + p_{2,i} \sum_{j=1}^J{\beta_{i}^{(j)} Logi \left( \mu_{\beta,i}^{(j)}, \sigma_{\beta, i}^{2(j)} \right)}  \right) \\
   & \left. \ast U \left(-\frac{1}{2},\frac{1}{2} \right) \right] (\hat{y}_i)
   = c \left(\hat{y}_i + \frac{1}{2} \right) - c \left( \hat{y}_i - \frac{1}{2} \right),
\end{split}
\end{equation}
where $i$ is the location index of the feature tensor, and $c(.)$ is the cumulative distribution function of the mixture model.


\subsection{Loss Function}

In this paper, we focus on optimizing the learned image compression to achieve the best rate-distortion (R-D) performance, based on the information theory. Let $R$ be the expected length of the bitstream, and $D$ be the reconstruction error between the source image and reconstructed image. The tradeoff between the rate and distortion is adjusted by a Lagrange multiplier denoted by $\lambda$. The objective cost function is then defined as follows:
\begin{equation}\label{total_loss}
\begin{aligned}
 L &=  \lambda D(\bm{x},\hat{\bm{x}})+H(\hat{\bm{y}})+H(\hat{\bm{z}}), \\
      H(\hat{\bm{y}}) &=  E [-\log_{2}(P_{\hat{\bm{y}}|\hat{\bm{y}}_c,\hat{\bm{z}}}(\hat{\bm{y}}|\hat{\bm{y}}_c,\hat{\bm{z}}))],\\
      H(\hat{\bm{z}}) &=  E [-\log_{2}(P_{\hat{\bm{z}}}(\hat{\bm{z}}))],
\end{aligned}
\end{equation}
where the distortion $D(\bm{x},\hat{\bm{x}})$ is the reconstruction error between origin image $\bm{x}$ and the decompressed image $\hat{\bm{x}}$. $H(\hat{\bm{y}})$ and $H(\hat{\bm{z}})$ are the entropies of the latents and hyperpriors, based on the estimated conditional probabilities, as a measure of the bits needed to encode them.

The mean square error (MSE) and MS-SSIM are the most widely used distortion metrics, which are chosen in this paper. Other terms such as the GAN cost can be added to the loss function to achieve other goals. However, when more terms are introduced, the optimized result has to trade off these different constraints.

\section{Experiment}
\label{Experiment}

\begin{figure*}[tb]
\centering
\subfigure[Original]{
\begin{minipage}[t]{0.25\linewidth}
\centering
\includegraphics[scale=0.40]{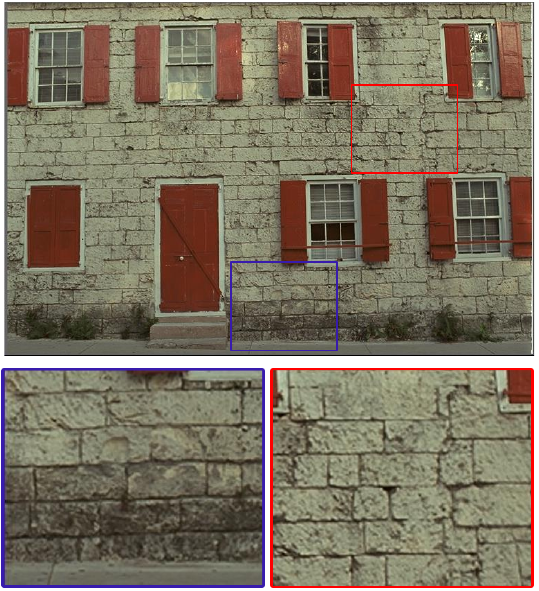}
\end{minipage}
}%
\subfigure[BPG (0.134/26.92/0.929)]{
\begin{minipage}[t]{0.25\linewidth}
\centering
\includegraphics[scale=0.40]{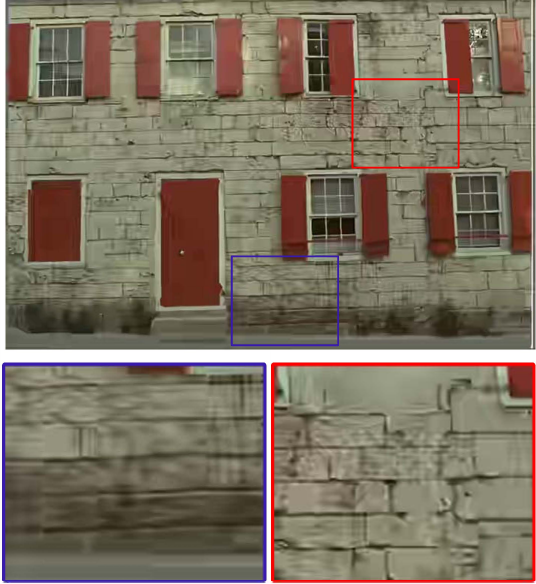}
\end{minipage}
}%
\subfigure[VVC (0.132/25.39/0.9039)]{
\begin{minipage}[t]{0.25\linewidth}
\centering
\includegraphics[scale=0.40]{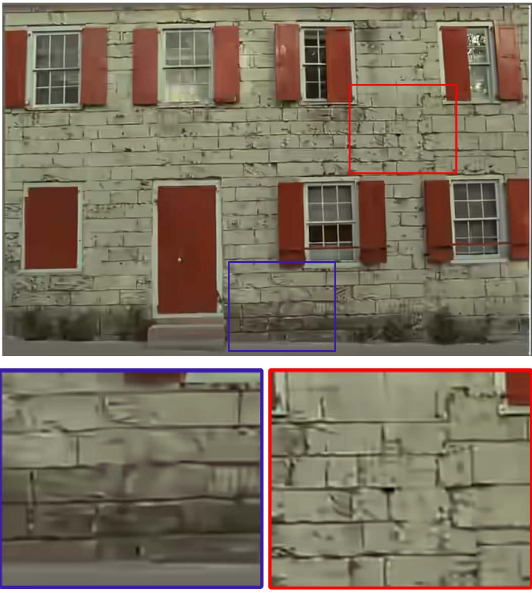}
\end{minipage}
}%
\subfigure[Ours (0.132/25.41/0.9133)]{
\begin{minipage}[t]{0.25\linewidth}
\centering
\includegraphics[scale=0.40]{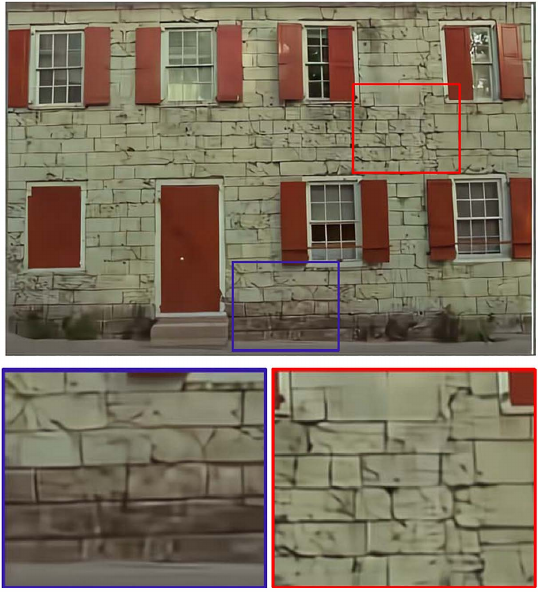}
\end{minipage}
}%
\centering
\caption{Example 1 in the Kodak dataset (bpp, PSNR (dB), MS-SSIM).}
\label{Example1}
\end{figure*}

In this section, we compare our method with different learning-based methods and traditional compression standards using the Kodak PhotoCD dataset \cite{Kodak} and two Tecnick datasets \cite{Tecnick}. The Kodak dataset consists of 24 images with a resolution of $768 \times 512$ or $512 \times 768$. The first Tecnick dataset named Tecnick-100 includes 100 uncompressed images with a resolution of $1200 \times 1200$. The second Tecnick dataset named Tecnick-40 has 40 uncompressed images with a resolution of $1200 \times 1200$.  The learning-based methods are recent state-of-the-art methods, including Balle2017 \cite{end_to_end}, Li2018 \cite{Limu_conf}, Lee2019 \cite{Lee_2020}, Cheng2020 \cite{cheng2020}, Lee2020 \cite{Lee_2021}, Li2020  \cite{Content_Weighted}, Chen2021 \cite{chen2021}, Ho2021 \cite{Ho2021}, Hu2021 \cite{Hu_2021}, Ma2022 \cite{Ma_2022_PAMI}, and  Zhu2022 \cite{Zhu_2022_CVPR}.  We also compare with traditional methods, including VVC-Intra (4:4:4) (version 12.1) \cite{VVC}, VVC-Intra (4:2:0),  BPG (4:4:4) \cite{BPG}, JPEG2000, WebP \cite{Webp}, and JPEG. Both PSNR and MS-SSIM metrics are used. The original MS-SSIM values are represented in dB scale by $-10 \log_{10}(1 - MS\textnormal{-}SSIM) $.

\subsection{Training Set}

The CLIC dataset \cite{CLIC} and LIU4K dataset \cite{LIU_dataset} are employed to train our models.  The images of the training dataset are rescaled to a resolution of $2000 \times 2000$, which is better for training. Data augmentation algorithms (i.e., rotation and scaling) are used to randomly crop 81,650 patches with a resolution of $384 \times 384$. The patches are stored as lossless PNG images.

\subsection{Parameter Settings}

We optimize the proposed models using mean square error (MSE) and MS-SSIM respectively. When optimized for MSE metric, the parameter $\lambda$ is selected from the set $\{0.0016,0.0032,0.0075,0.015, 0.023, 0.03, 0.045\}$. Each value trains a network for a particular bit rate. The number of filters $N$ is set to 128 for the three lower bit rates, and is set to 256 for the four higher bit rates. When the MS-SSIM metric is targeted, the parameter $\lambda$ is in the set $\{12, 40, 80, 120\}$.  The number of filters $N$ is set to 128 for the two lower bit rate, and is set to 256 for the two higher bit rates. Each model was trained up to $1.5 \times 10^{6}$ iterations to obtain stable performance. The Adam solver with a batch size of 8 is adopted. The learning rate is set to $1 \times 10^{-4}$ in the first 750,000 iterations, and we gradually reduce the learning rate by 0.5 after every 100,000 iterations in the last 750,000 iterations.

\subsection{Performances on Kodak, Tecnick-100, and Tecnick-40 Datasets}

The average MS-SSIM and PSNR performances over the 24 Kodak images are illustrated in Fig. \ref{test_kodak}. Note that in Fig. \ref{test_kodak} to Fig. \ref{test_Tecnick40}, the notations \textit{Scheme (PSNR)} and \textit{Scheme (MS-SSIM)} in the legends mean that the model in the scheme is optimized for PSNR and MS-SSIM respectively.

In Fig. \ref{test_kodak}, when optimized for PSNR, Lee2020 (PSNR) \cite{Lee_2021} is the best among previous methods, which obtains even better performance than VVC (4:4:4) at high rates. The next closest method to VVC (4:4:4) is Cheng2020 \cite{cheng2020}. When the bit rate is less than 0.3 bpp, our method has similar performance to \cite{Lee_2021} and VVC (4:4:4). When the bit rate is higher than 0.4 bpp, our method achieves the best performance, which is 0.2-0.3 dB over \cite{Lee_2021} and 0.3-0.4 dB over VVC (4:4:4).

In the MS-SSIM results in Fig. \ref{kodak_SSIM},  Lee2020 (MS-SSIM) \cite{Lee_2021} also achieves  better performance than previous learning-based methods and all the traditional image codecs including VVC (4:4:4).  Our proposed method optimized for MS-SSIM achieves slightly better results than Lee2020 (MS-SSIM).

Fig. \ref{test_Tecnick} compares the performances of different methods on the Tecnick-100 dataset. Our scheme also outperforms all available learning-based methods and all the traditional image codecs including VVC (4:4:4) in term of both PSNR and MS-SSIM. Our method is the only method better than VVC. When the bit rate is above 0.2bpp, our method is about 0.2-0.3 dB higher than VVC (4:4:4).

Fig. \ref{test_Tecnick40} compares the performances of some methods on the Tecnick-40 dataset. The results of other methods are not available. Our scheme also outperforms other learned methods including Lee2020 \cite{Lee_2021} and VVC (4:4:4) in both PSNR and MS-SSIM.

One example are shown in Fig. \ref{Example1} to compare the visual quality of different methods. Our method produces the most visually pleasing results.

\subsection{Ablation Studies}
\label{Ablation}

\begin{figure}[tb]
	\centering
		\includegraphics[scale=0.6]{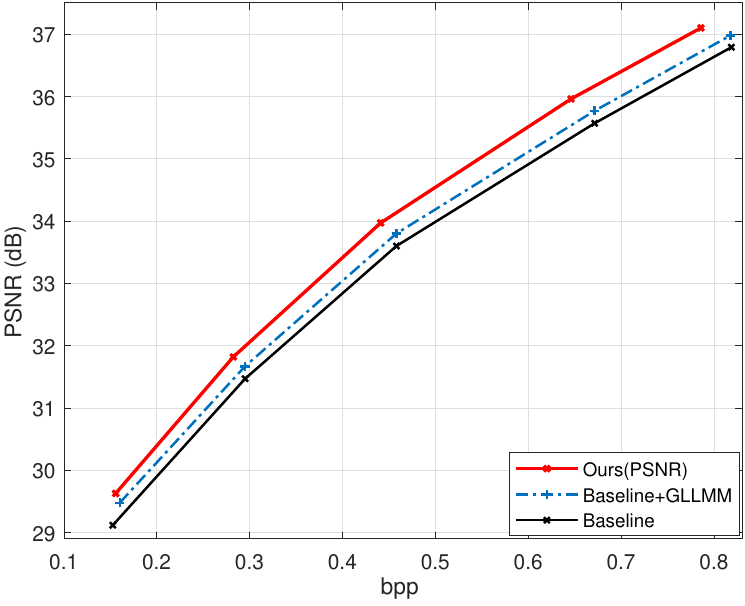}
	\caption{Contributions of GLLMM and CRM on the Kodak dataset. The scheme in \cite{cheng2020} is the baseline.}
	\label{fig_CRM_GLLMM}
\end{figure}

\subsubsection{Contributions of CRM and GLLMM}

We first present ablation study to show the gain of the concatenated residual module (CRM) and the GLLMM model. The scheme in \cite{cheng2020} is used as the baseline.  On top of the baseline, we add different modules in turn. In order to compare as fair as possible, the parameter $\lambda$ is chosen in the set $\{0.0032,0.0075, 0.015, 0.03, 0.045\}$. The number of filters $N$ is set to 128 for the two lower bit rates, and 256 for the three higher bit rates. Other training setups are the same. The MSE objective function is optimized in the ablation experiments.


The results are shown in Fig. \ref{fig_CRM_GLLMM}.  We first replace the GMM in the baseline by GLLMM, denoted as Baseline+GLLMM, which improves the R-D  performance by about 0.15 dB at the same bit rate. Compared to Baseline+GLLMM, the proposed full method in this paper has a further improvement of 0.3 dB, which is the contribution of the CRM over the original residual blocks.

\subsubsection{Number of Concatenated Residual Modules}

\begin{figure}[tb]
	\centering
    \includegraphics[scale=0.55]{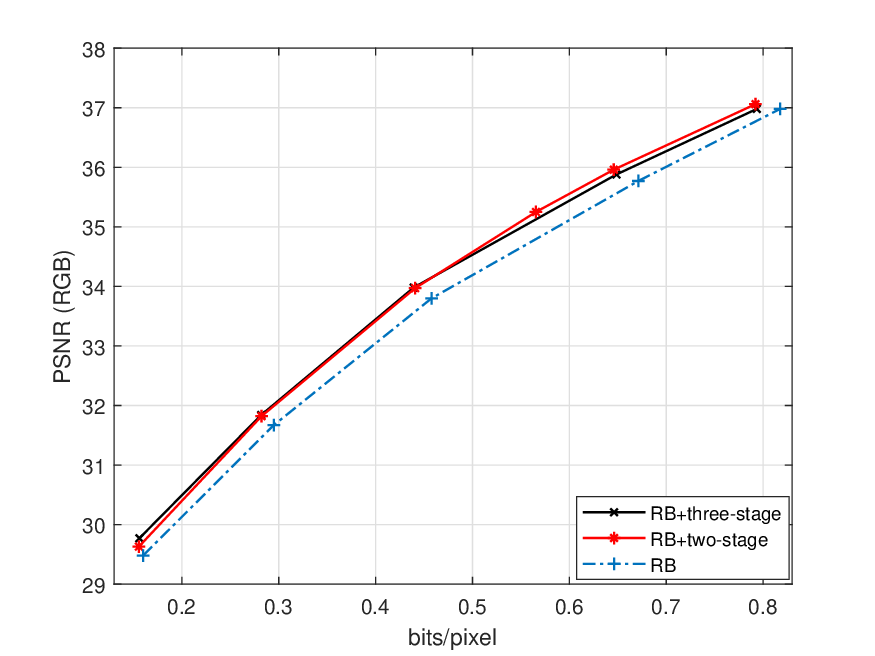}
	\caption{The impact of concatenated residual modules on the Kodak dataset.}
	\label{abalation_residual_block}
\end{figure}

Next, we compare the standard residual block (RB), two-stage concatenated residual module (RB+two-stage), and three-stage concatenated residual module (RB+three-stage). The results are shown in Fig. \ref{abalation_residual_block}.

The method with RB \cite{resblock} achieves the worst performance. RB+two-stage method is 0.3 dB higher than RB. The RB+three-stage achieves the same performance with RB+two-stage at low bit rates and is sightly worse at high bit rates. Moreover, the model size will increase about $13\%$. Therefore, we adopt the RB+two-stage method in our framework.

\subsubsection{Comparisons of Different Entropy Coding Models}

\begin{figure}[tb]
	\centering
		\includegraphics[scale=0.6]{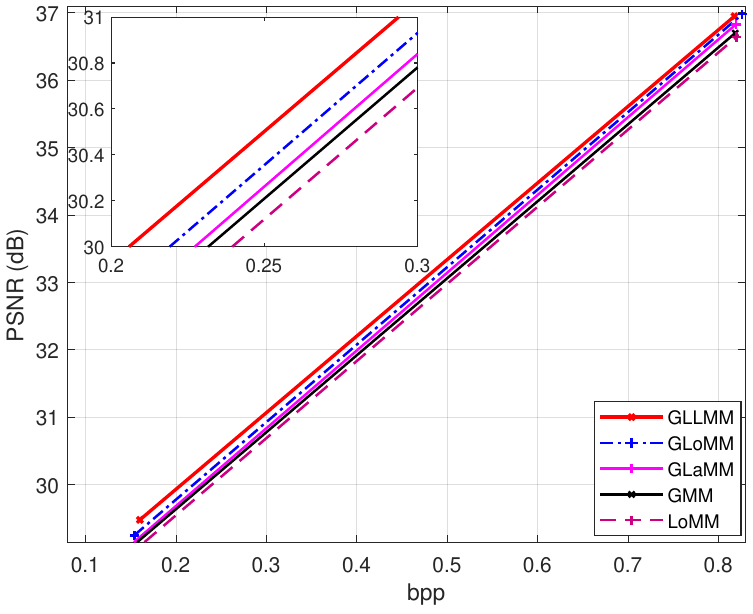}
	\caption{The comparisons of different entropy coding models on the Kodak dataset.}
	\label{different_PDF}
\end{figure}

In Fig. \ref{different_PDF}, we use the Kodak dataset to compare the performances of different entropy coding models, including Logistic mixture model (LoMM), Gaussian mixture model (GMM), Gaussian-Logistic mixture model (GLoMM), Gaussian-Laplacian mixture model (GLaMM), and the proposed Gaussian-Logistic-Laplacian mixture model (GLLMM). If some distributions are not used, we set the corresponding $K$, $M$ or $J$ in Eq. (\ref{GLLMM_Eq}) to be $0$. It can be seen that LoMM achieves the worst performance. GMM is slightly better than LoMM, which agrees with the results in \cite{cheng2020}. Adding either Laplacian or Logistic distribution to GMM can further improve its performance, with more gains from Logistic than Laplacian. Finally, adding both Laplacian and Logistic distributions as in the proposed GLLMM achieves the best result. The gaps between different curves are quite consistent at all bit rates.

\begin{figure*}[tb]
	\flushleft
		\includegraphics[scale=0.37]{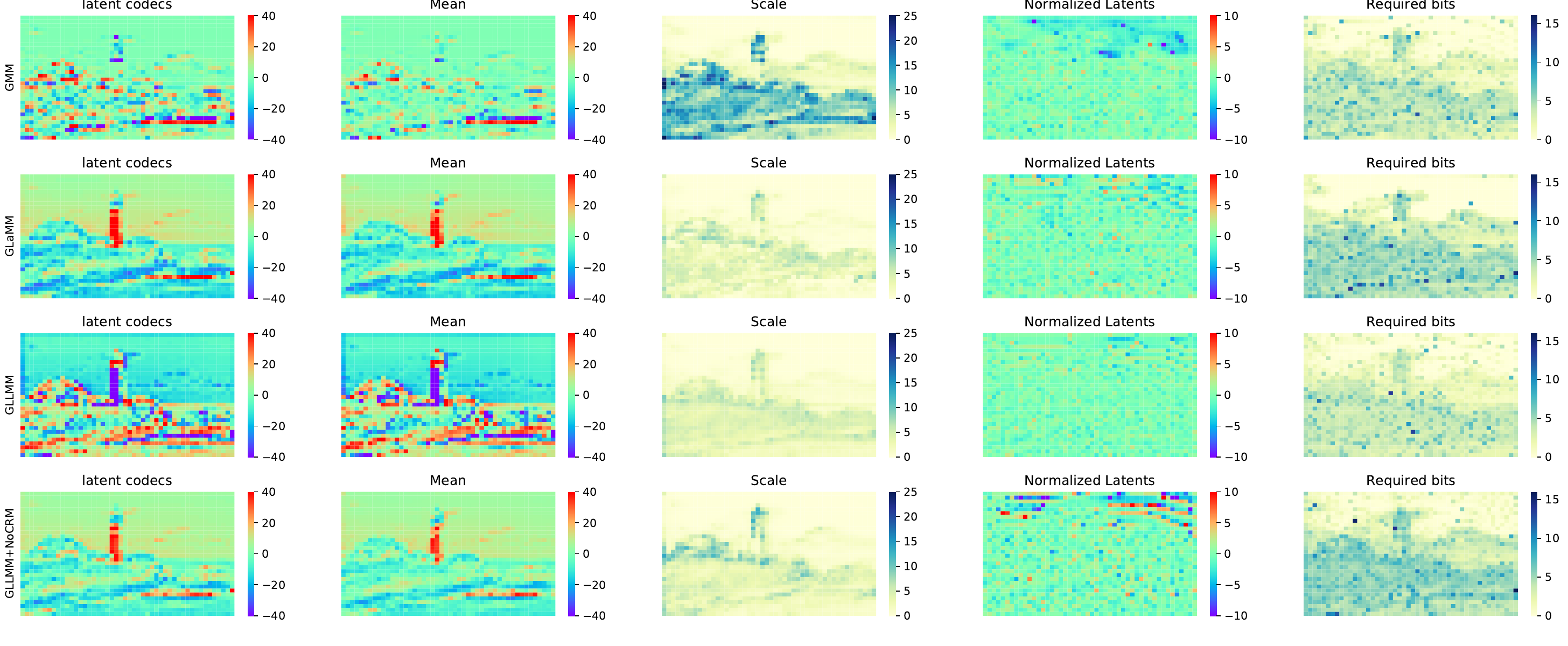}
	\caption{Visualization of different models for the feature map with the highest entropy using the Kodim21 image from Kodak dataset, where NoCRM means that the CRM modules are removed from the proposed method.}
	\label{entropy_model}
\end{figure*}

In Fig. \ref{entropy_model}, we use image Kodim21 from the Kodak dataset to visualize the impacts of different entropy models using the same network architecture. The feature map with the highest entropy is displayed. The first column shows the values of the quantized latents. The second and third columns are the predicted parameters $\bm{\mu}$ and $\bm{\sigma}$. For the proposed GLLMM model, the mean is predicted by
\begin{equation}\label{GLLMM_Eq_mean}
   \mu_i = p_{0,i} \sum_{k=1}^K{w_{i}^{(k)}\mu_{w,i}^{(k)}} +p_{1,i} \sum_{m=1}^M{\alpha_{i}^{(m)}\mu_{\alpha,i}^{(m)}} + p_{2,i}\sum_{j=1}^J{\beta_{i}^{(j)}\mu_{\beta,i}^{(j)}.}
\end{equation}
The means and variances for other models are obtained similarly. If some distributions are not used, the corresponding $K$, $M$ or $J$ are set to $0$.

The fourth column in Fig. \ref{entropy_model} shows the normalized latents $\frac{\hat{y_i}-\mu_i}{\sigma_i}$, as used in \cite{Joint,cheng2020,chen2021}, because if the model is accurate, the normalized result should have zero mean and unit variance. Therefore the normalized result helps to visualize the remaining redundancy that is not captured by the entropy models. The last column shows the required bits to encoder the latents at each position, which is calculated using the entropy of the predicted discretized distribution at that position.

Table \ref{calculate_mu_variance} reports the mean and variance of the images in the fourth column and the average bits of the images in the fifth column.

It can be observed from Fig. \ref{entropy_model} and Table \ref{calculate_mu_variance} that our GLLMM model provides more uniform normalized latents or less remaining redundancy, and needs less bits for encoding.

\begin{figure*}[tb]
\centering
\subfigure[bpp:0.97,PSNR:37.60]{
\begin{minipage}[t]{0.25\linewidth}
\centering
\includegraphics[scale=0.12]{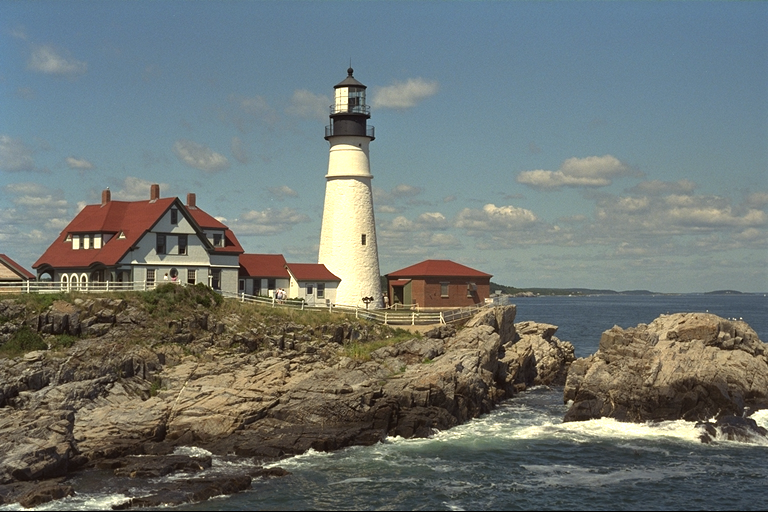}
\end{minipage}
}%
\subfigure[Latent codecs]{
\begin{minipage}[t]{0.25\linewidth}
\centering
\includegraphics[scale=0.6]{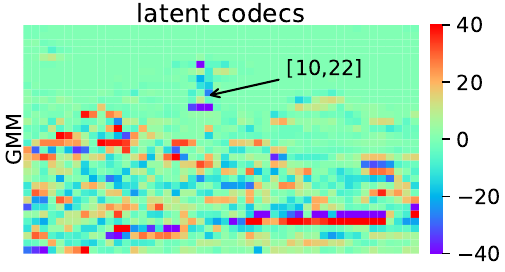}
\end{minipage}
}%
\subfigure[Probability Distribution Function]{
\begin{minipage}[t]{0.25\linewidth}
\flushright
\includegraphics[scale=0.3]{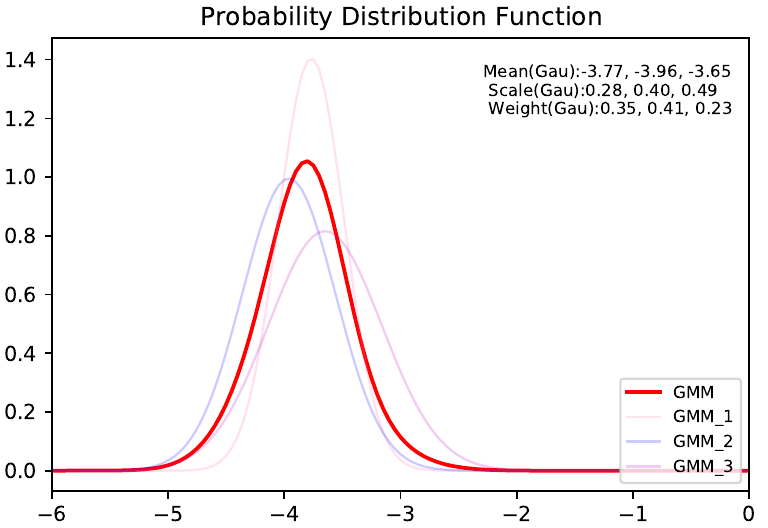}
\end{minipage}
}%
\subfigure[Histogram]{
\begin{minipage}[t]{0.25\linewidth}
\flushleft
\includegraphics[scale=0.3]{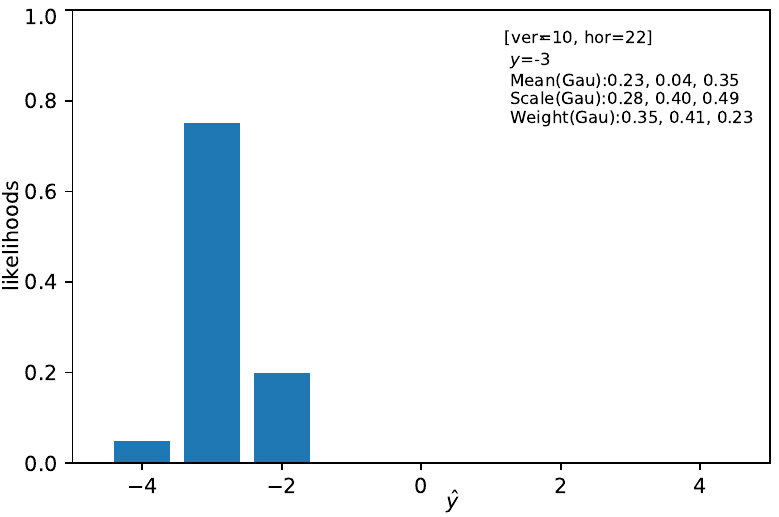}
\end{minipage}
}%

\subfigure[bpp:0.97,PSNR:37.71]{
\begin{minipage}[t]{0.25\linewidth}
\centering
\includegraphics[scale=0.12]{./test_result/histogram/keda_21.origin.png}
\end{minipage}
}%
\subfigure[Latent codecs]{
\begin{minipage}[t]{0.25\linewidth}
\centering
\includegraphics[scale=0.6]{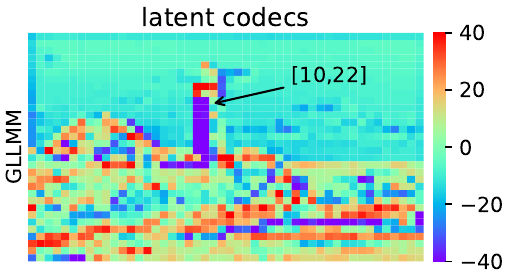}
\end{minipage}
}%
\subfigure[Probability Distribution Function]{
\begin{minipage}[t]{0.25\linewidth}
\flushright
\includegraphics[scale=0.3]{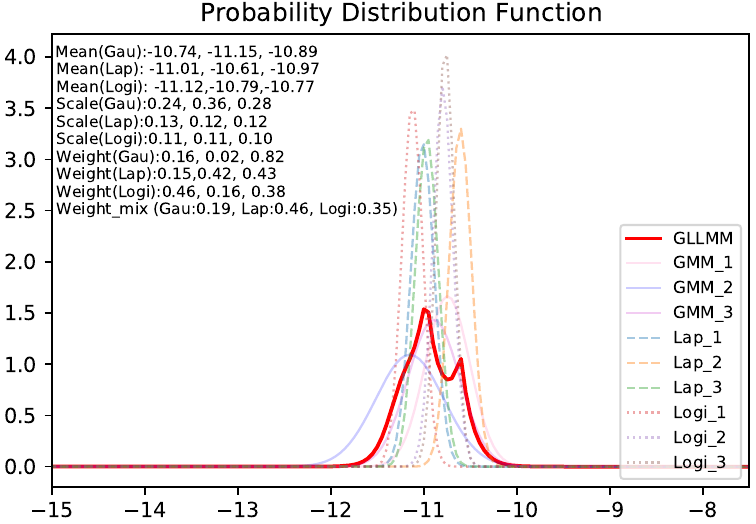}
\end{minipage}
}%
\subfigure[Histogram]{
\begin{minipage}[t]{0.25\linewidth}
\flushleft
\includegraphics[scale=0.3]{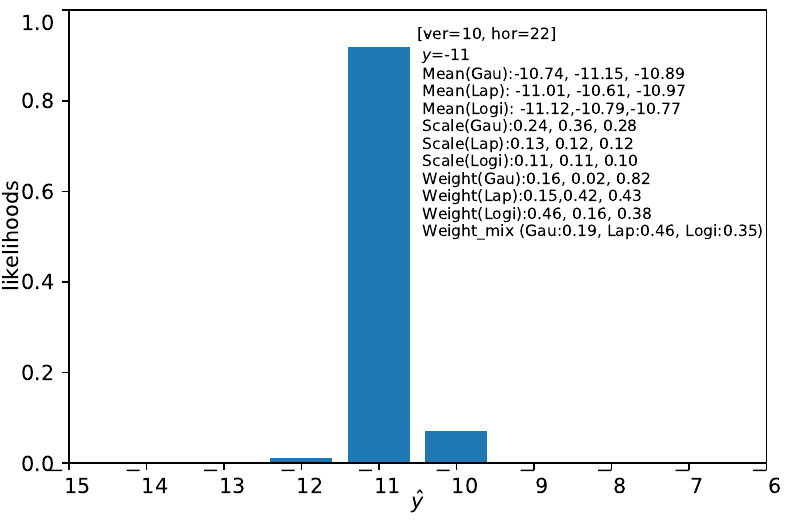}
\end{minipage}
}%

\centering
\caption{Visualization of the estimated probabilities at location [10, 22] of the feature map in Fig. \ref{entropy_model} using GMM and GLLMM.}
\label{entropy_histograml}
\end{figure*}

Fig. \ref{entropy_histograml} further zooms in to one particular location at coordinate [10, 22] in the feature map in Fig. \ref{entropy_model}, and plots the estimated continuous and quantized distributions using GMM and GLLMM respectively. The first column shows the reconstructed images of the two methods. The second column indicates the location of the selected latent. The third column plots the estimated individual distributions, the final mixed probability, and the corresponding parameters. It can be seen that in GLLMM, all the components in it are used effectively to represent this complex probability. The last column shows the histograms of the quantized probabilities.

\begin{table}[h]
\caption{Numerical results for Columns 4 and 5 in Fig. \ref{entropy_model}.}
\begin{center}
\begin{tabular}{ccccc}
\hline
\textbf{Method} & \textbf{Col. 4} & \textbf{Col. 4} &\textbf{Col. 5} \\
     & \textbf{Mean} & \textbf{Variance} &\textbf{Mean} \\
\hline
GMM    &0.390 & 2.214 & 4.202 \\
GLaMM  &0.077 & 2.412 & 6.314 \\
GLLMM  &0.049  &1.303 & 3.359 \\
GLLMM+NoCRM  &0.060 & 4.773 & 8.386 \\

\hline
\end{tabular}
\end{center}
\label{calculate_mu_variance}
\end{table}


\begin{figure}[!htb]
	\centering
     \includegraphics[scale=0.4]{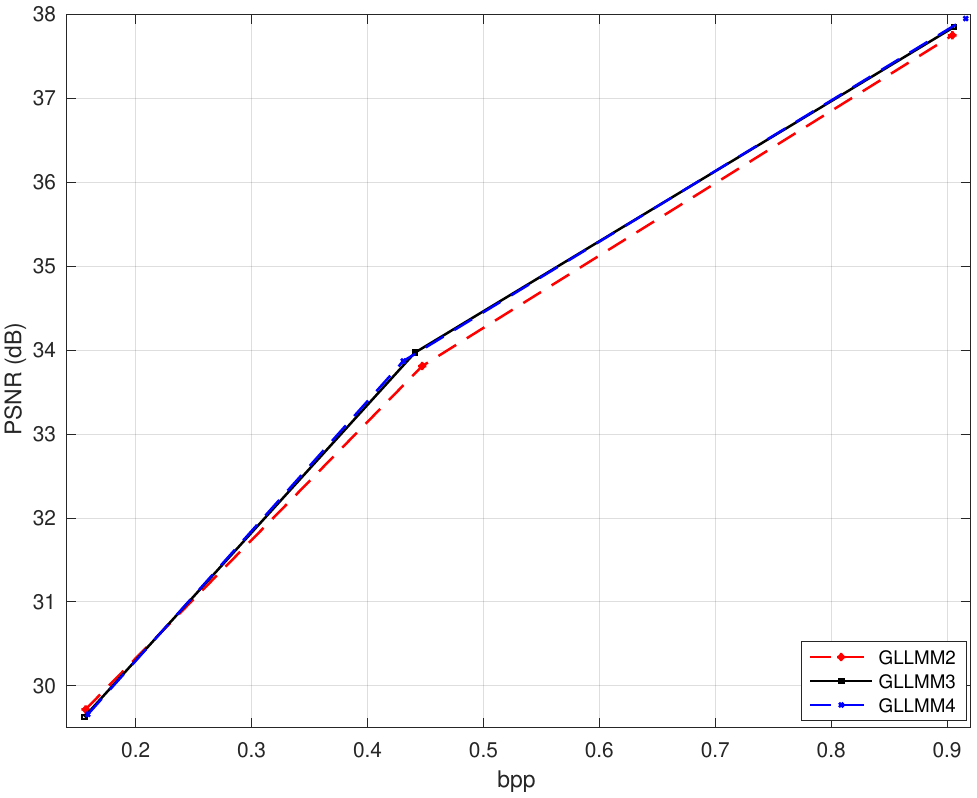}
	\caption{The performance of GLLMM with different orders on the Kodak dataset. GLLMMn means that $K=M=J=n$ in Eq. (\ref{GLLMM_Eq}).}
	\label{different_combination}
\end{figure}

\subsubsection{GLLMM with Different Orders}

We next compare the performances of GLLMM with different orders. We use GLLMMn to represent the GLLMM with $K=M=J=n$ in Eq. (\ref{GLLMM_Eq}). The results are shown in Fig. \ref{different_combination}. It can be seen that GLLMM2 achieves the worst performance. GLLMM3 is up to about $0.15-0.2$ dB higher than GLLMM2. GLLMM4 almost achieves the same performance as GLLMM3. Therefore, the values of K, M, and J in other experiments of this paper are set to be $3$.

\begin{figure}[!htb]
	\centering
		\includegraphics[scale=0.43]{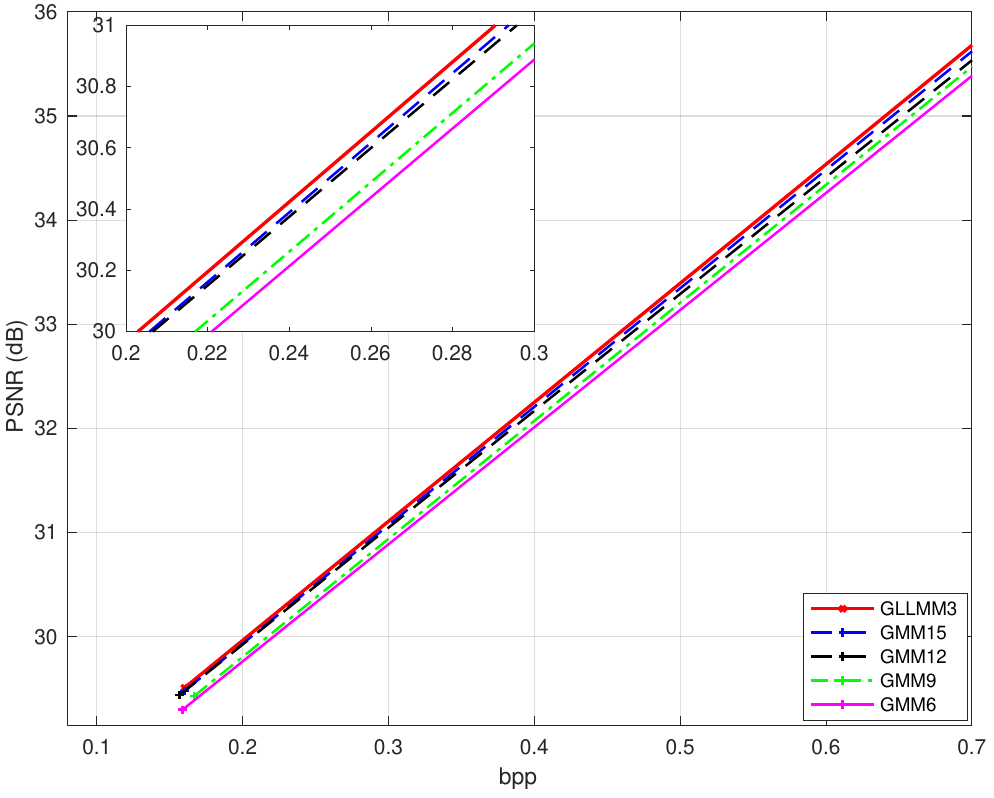}
	\caption{Comparison of GLLMM3 with different orders of GMM on the Kodak dataset.}
	\label{Different_GMMs}
\end{figure}

\subsubsection{Comparison of GLLMM and Higher-order GMM}

We also compare the performance of GLLMM3 and higher-order GMM, denoted as GMMn. Theoretically, GMM with enough orders can achieve any distribution. However, this comes at the price of increased complexity, and the gain diminishes gradually. For each method, we train two models at low and high bit rates respectively, with $N=128$. The results are shown in Fig. \ref{Different_GMMs}. It can be seen that higher-order GMM has better performance. However, the performance saturates gradually. GLLMM3 has better performance than GMM15. Moreover, GLLMM3 has 30 parameters in Eq. (\ref{GLLMM_Eq}), whereas GMM15 has 45 parameters in Eq. \ref{GMM_Eq}. As a result, compared to GLLMM3, the encoding and decoding time of GMM15 is 52.32\% slower, its model size is 17.07\% larger, and training time is 8.23 \% longer. Therefore the proposed GLLMM is more effective than GMM when complexity is considered.

\subsubsection{Impact of the Attention Module and Context Model}

\begin{figure}[!htb]
   \centering
    \includegraphics[scale=0.43]{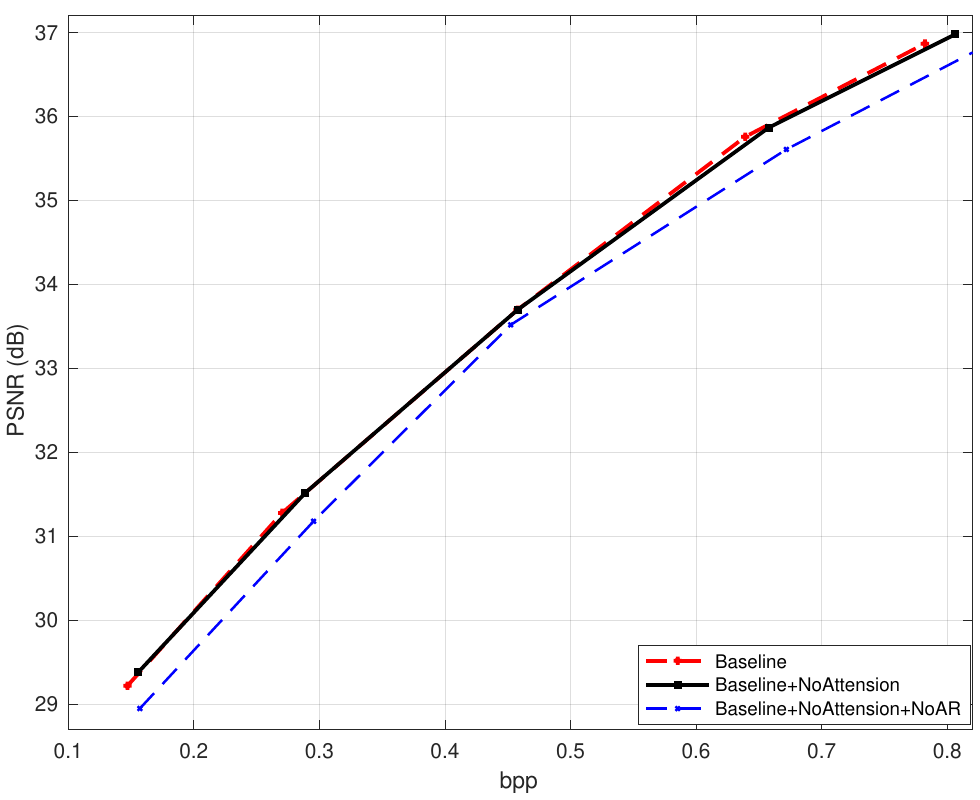}
	\caption{The impact of attention module and context model on the Kodak dataset.}
	\label{fig_attention_context}
\end{figure}

Fig. \ref{fig_attention_context} shows the effectiveness of the attention module and context model. In this figure, the baseline is the proposed method with CRM and GLLMM, which achieves the best performance. In Baseline+NoAttension, the attention module is removed, which has similar performance to the Baseline. In Baseline+NoAttension+NoAR, the autoregressive (AR) context model is also removed, which is 0.2-0.3 dB lower than the Baseline.

\subsubsection{Impact of Different Loss Functions}

\begin{figure}[!htb]
    \centering
    \includegraphics[scale=0.4]{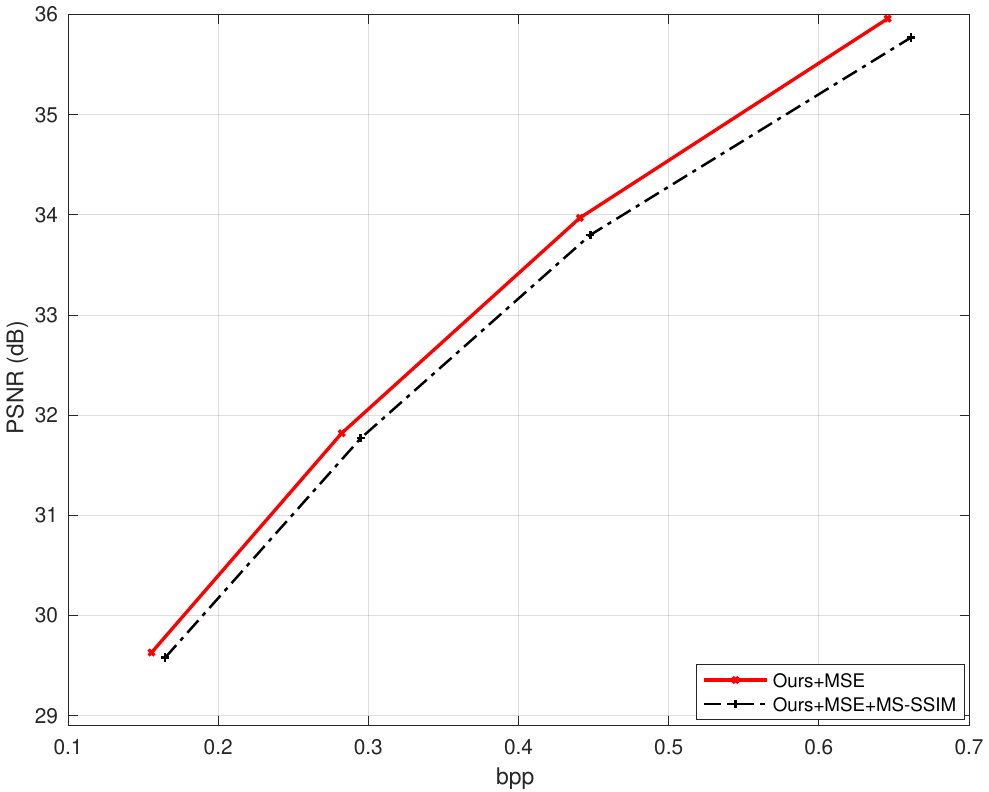}
	\caption{The impact of different loss functions on the Kodak dataset. }
	\label{fig_differentLossFunction}
\end{figure}

Fig. \ref{fig_differentLossFunction} shows the impact of different loss functions. In this figure, the Ours+MSE is the proposed method with only MSE as the distortion measure $D(\bm{x},\hat{\bm{x}})$ in the loss function, which achieves the best PSNR performance. In Ours+MSE+MS-SSIM, we include both the MSE and MS-SSIM in the distortion part of the loss function, whose PSNR is 0.2-0.3 dB lower than the loss function with MSE-only distortion.

\subsubsection{Impact of the Post-Processing Module in \cite{Lee_2021}}
We also study the effectiveness of the post-processing module on image compression performance in our method. In \cite{Lee_2021}, the post-processing module is combined to the image compression framework, which achieves 0.5 dB gain compared to its baseline without the post-processing. The post processing can be applied to other learning-based schemes as well. To study its effectiveness in our scheme,  we implement the same post-processing as in \cite{Lee_2021} into our method. The corresponding results are listed in Table \ref{LR_HR}. We can observe that these two schemes have almost the same performances on Kodak dataset. The main reason is that we have much better residual modules and entropy model. Therefore there is no need to apply the post processing. This also reduces the complexity of the scheme.
\begin{table*}[!thp]
\caption{Impact of the Post-Processing Module in \cite{Lee_2021} for Kodak Dataset}
\begin{center}
  \begin{tabular}{ccccccc}
  \hline
   \textbf{Method}  &  \textbf{Number of Filters}& \textbf{$\lambda$}& \textbf{Objective Function} & \textbf{BPP}& \textbf{PSNR} &\textbf{Ms$-$SSIM} \\
  \hline
  \textbf{Ours} &128&0.0032 &PSNR &0.1556&29.63 dB &12.59 dB\\
  \textbf{Ours+post-processing} &128&0.0032 &PSNR &0.1554&29.65 dB &12.59 dB\\
  \textbf{ours} &256 &0.015 &PSNR &0.4408 &33.97 dB &16.70 dB\\
  \textbf{Ours+post-processing} &256&0.015 &PSNR &0.4400 &33.98 dB &16.71 dB\\
  \hline
\end{tabular}
\label{LR_HR}
\end{center}
\end{table*}

\subsection{Encoding and Decoding Complexity}

\begin{table*}[!thp]
\caption{Comparisons of encoding/decoding time and model sizes for Kodak Dataset.}
\begin{center}
\begin{tabular}{ccccc}
\hline
\textbf{Method} & \textbf{Encoding Time} & \textbf{Decoding Time} &\textbf{Model Size (Low Rates) }&\textbf{Model Size (High Rates)} \\
\hline
VVC \cite{VVC} & 402.27s& 0.61s& 7.2 MB & 7.2MB\\
Lee2019\cite{Lee_2020}&15.721s& 42.88s& 123.8 MB & 292.6MB\\
Hu2020 \cite{Hu_2021}  &281.25s& 450.23s& 84.6 MB & 290.9MB\\
Cheng2020 \cite{cheng2020} &20.89s& 22.14s& 50.8 MB & 175.18MB\\
Chen2021 \cite{chen2021} &402.26s& 2405.14s& 200.99 MB & 200.99MB\\
GLLMM     &385.26s& 387.62s& 77.08 MB & 241.03MB\\
\hline
\end{tabular}
\end{center}
\label{runing_time}
\end{table*}

Table \ref{runing_time} compares the complexities of different approaches. Since VVC, Hu2020 \cite{Hu_AAAI} and Cheng2020 \cite{cheng2020} only run on CPU, we evaluate the encoding and decoding time of different methods at the similar bit rate on an 2.9GHz Intel Xeon Gold 6226R CPU. The average time over all Kodak images is used. The average model sizes at low bit rates and high bit rates are also reported.

It can be seen from Table \ref{runing_time} that compared to Chen2021 \cite{chen2021}, the proposed scheme is faster in both encoding and decoding, but is slower than other methods. Our model size is also smaller than \cite{chen2021}. Compared to VVC, our encoding is faster, but the decoding is much slower.

%
%

\section{Conclusions}
\label{Conclusion}


In this paper, we improve the state of the art of learning-based image compression by presenting a more flexible conditional probability model based on the discretized Gaussian-Laplacian-Logistic mixture distribution, which captures the spatial-channel correlation more effectively in latent representations. We also develop an improved concatenated residual block module for the encoder network.

Experiments demonstrate that the proposed method outperforms VVC (4:4:4) in terms of both PSNR and MS-SSIM metrics when measured on Kodak, Tecnick-100, and Tecnick-40 dataset. Also, our scheme achieves better performance compared to all the previous state-of-the-art learning-based methods.

Our method still has some rooms to improve. Since the autoregressive context model is employed in our framework, the symbols have to be decoded in an serial manner. Although it can effectively reduce the spatial correlation of the latent representations, it significantly increases the time complexity. How to reduce the complexity of the context model without too much degradation of the performance is a future research topic. Also, the complexity of our method can be further optimized by different approaches such as model compression and optimization.

Another possible future topic is to develop low-cost multivariate mixture models for learned image compression.

%

\ifCLASSOPTIONcaptionsoff
  \newpage
\fi


\begin{IEEEbiography}[{\includegraphics[width=1in,height=1.25in,clip,keepaspectratio]{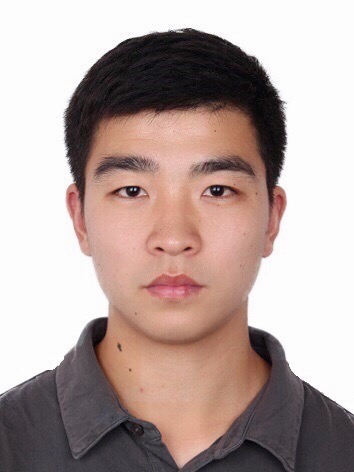}}]{Haisheng Fu} (Student Member, IEEE) received the B.S. degree in automation engineering from Henan Polytechnic University, China. He is currently pursuing the Ph.D. degree in electronic science and technology with Xi’an Jiaotong University, Xi’an. He is currently a visiting student of Simon Fraser University, Canada. His research interests include Machine Learning, Image and Video compression, Deep Learning, and VLSI design.

\end{IEEEbiography}

\begin{IEEEbiography}[{\includegraphics[width=1in,height=1.25in,clip,keepaspectratio]{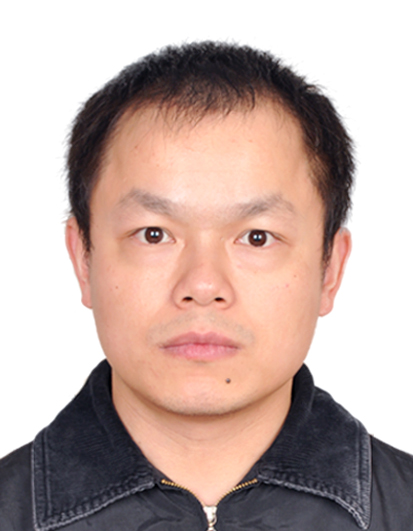}}]{Feng Liang} is currently Professor of the Microelectronics School at Xi'an Jiaotong University. He earned his B.E. from Zhengzhou University and his M.E. and Ph.D. from Xi'an Jiaotong University. His current research interests include Signal Processing, Machine Learning, VLSI design, CIM, and computer architecture.
\end{IEEEbiography}

\begin{IEEEbiography}[{\includegraphics[width=1in,height=1.25in,clip,keepaspectratio]{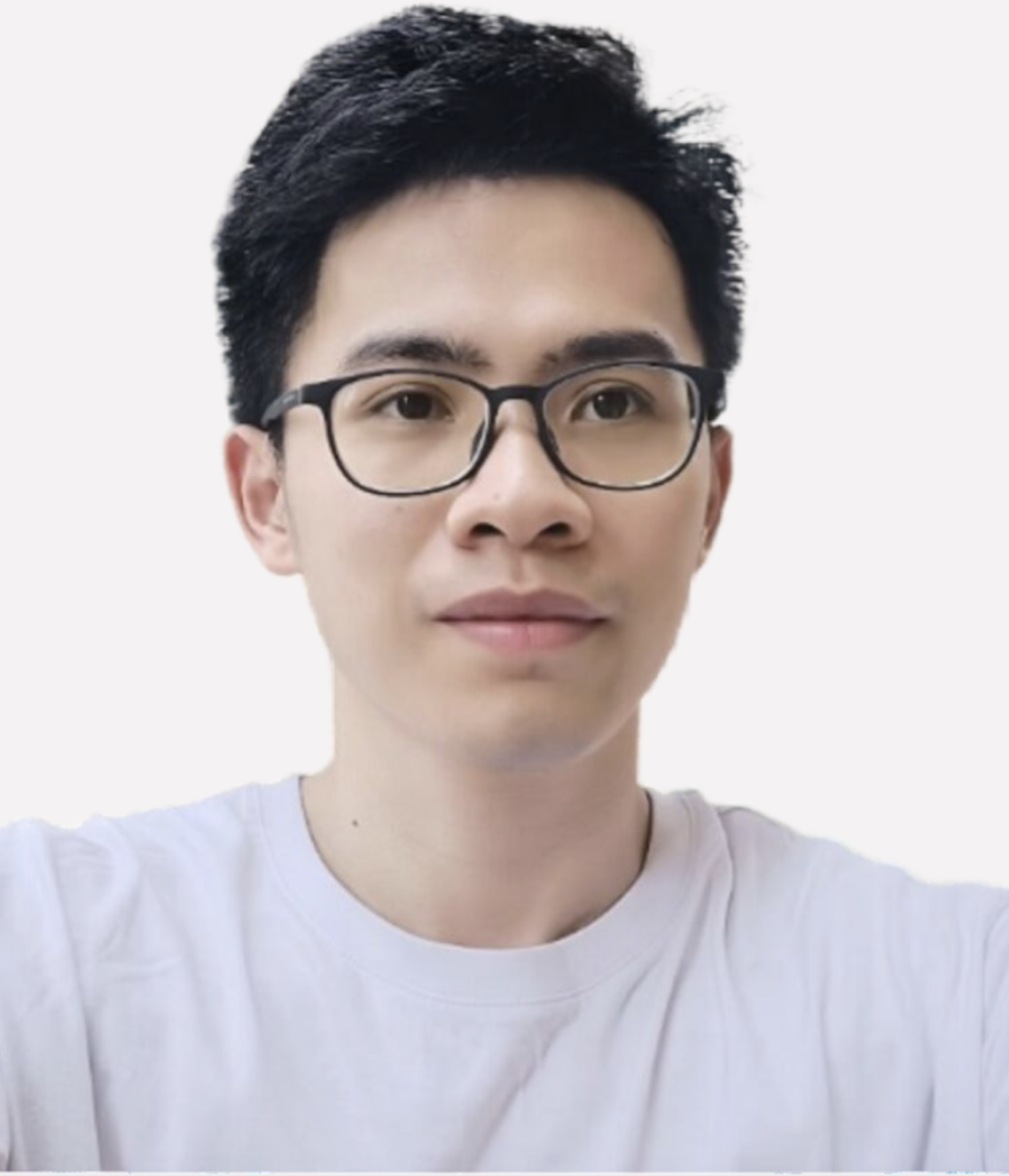}}]{Jianping Lin}  received the B.S. degree in electronic engineering from the University of Science and Technology of China, Hefei, China, in 2016, where he is currently pursuing the Ph.D. degree with the Department of Electronic Engineering and Information Science. His research interests mainly include video coding/processing and machine learning
\end{IEEEbiography}

\begin{IEEEbiography}[{\includegraphics[width=1in,height=1.25in,clip,keepaspectratio]{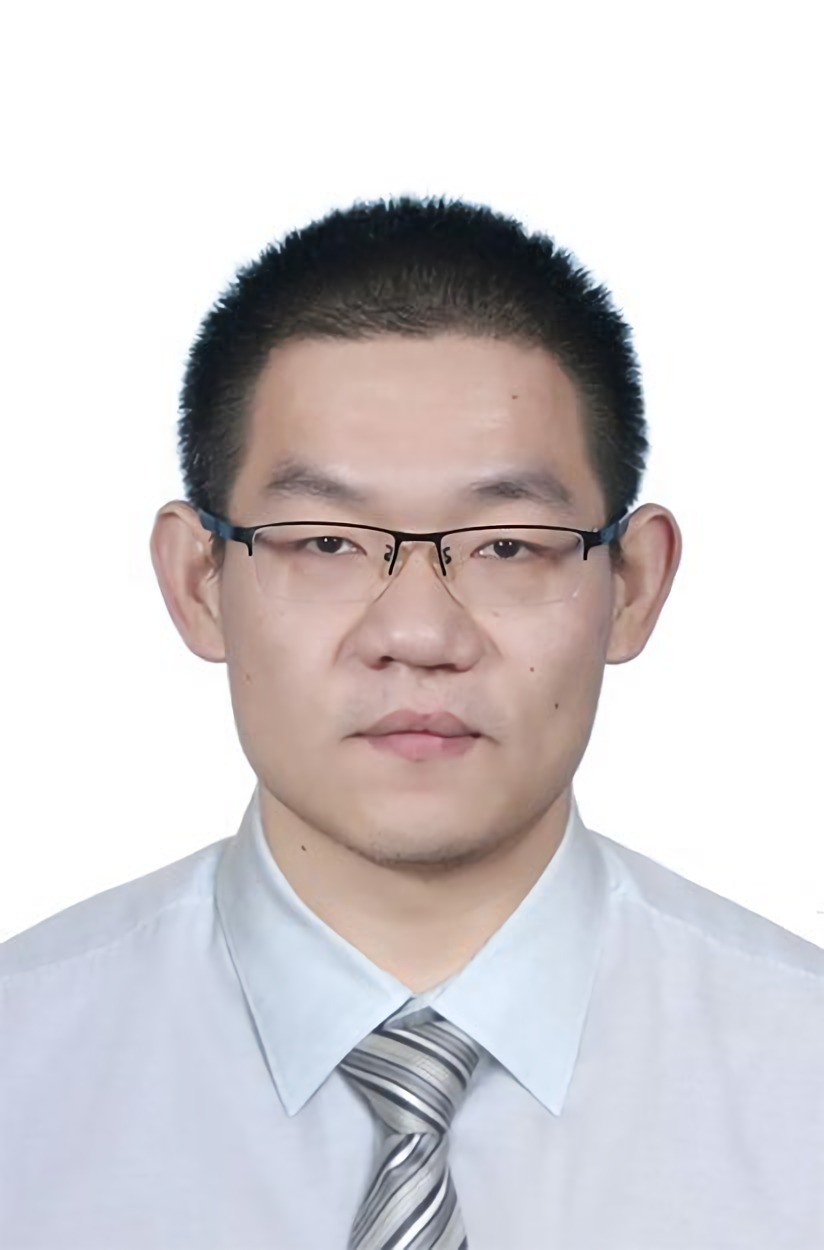}}]{Bing Li } received his Ph.D. degree from Xi’an Jiaotong University, Xi’an, in 2021. He is now an engineer in Huawei. His current research interests include Elliptic curve cryptosystem, machine learning, and hardware implementation of neural networks.
\end{IEEEbiography}

\begin{IEEEbiography}[{\includegraphics[width=1in,height=1.25in,clip,keepaspectratio]{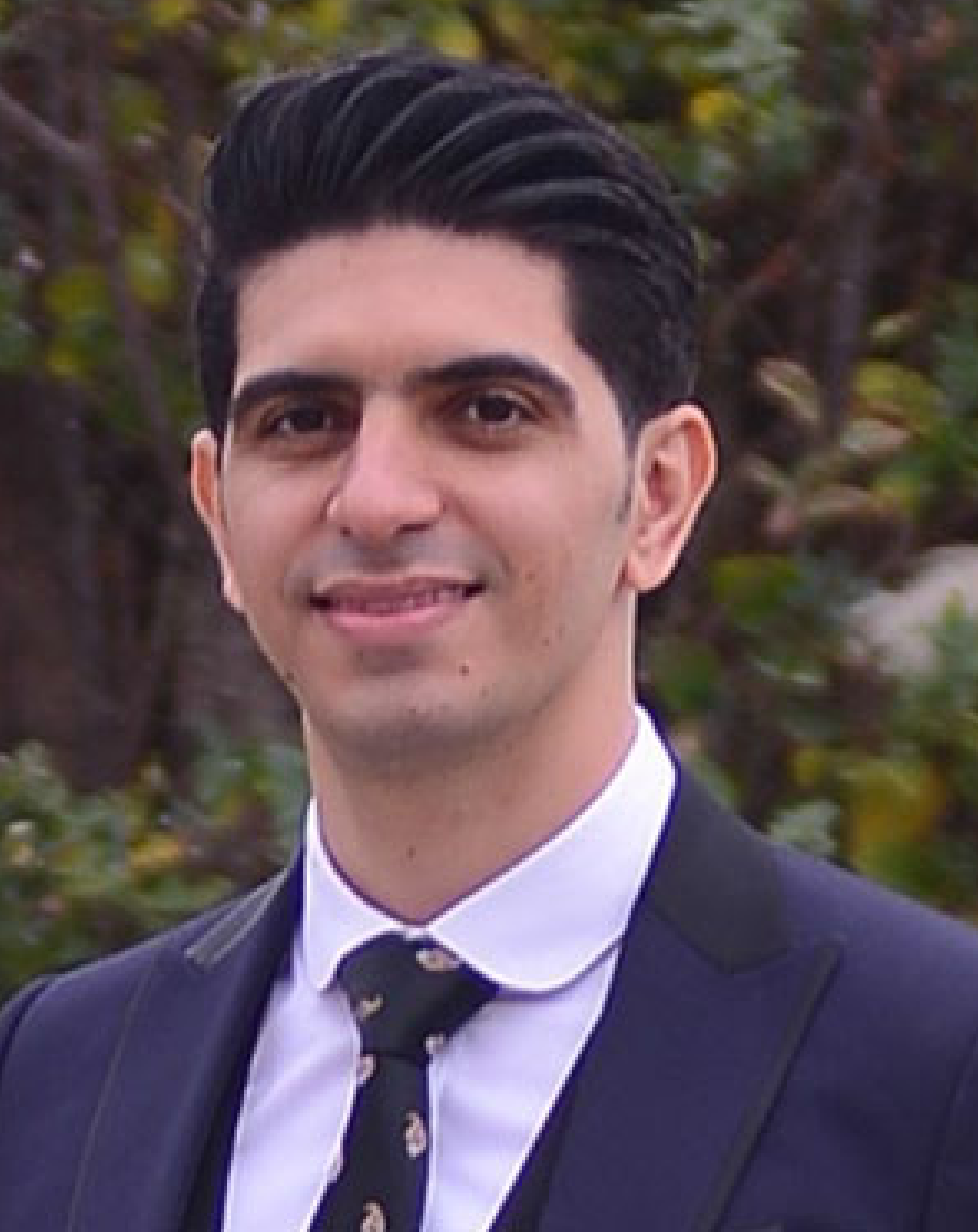}}]{Mohammad Akbari} (Student Member, IEEE) received the B.S. degree in software engineering from
the Shahid Bahonar University of Shiraz, Shiraz,
Iran, in 2010, the M.Sc. degree in computer science
from the University of Lethbridge, Lethbridge, AB,
Canada, in 2014, and the Ph.D. degree in engineering
science from Simon Fraser University, Burnaby, BC,
Canada, in 2020. He is currently an AI Researcher
with Huawei Technologies, Markham, ON, Canada.
His research interests include deep learning, learned
image compression, and music information retrieval.
He was the recipient of the Best Student Paper Award finalist at the 2020 International Conference on Multimedia and Expo, the 2015 Convocation Medal
of Merit from University of Lethbridge, and the Winner of the 2014 Canadian
Microsoft Imagine Cup Innovation Competition.
\end{IEEEbiography}

\begin{IEEEbiography}[{\includegraphics[width=1in,height=1.25in,clip,keepaspectratio]{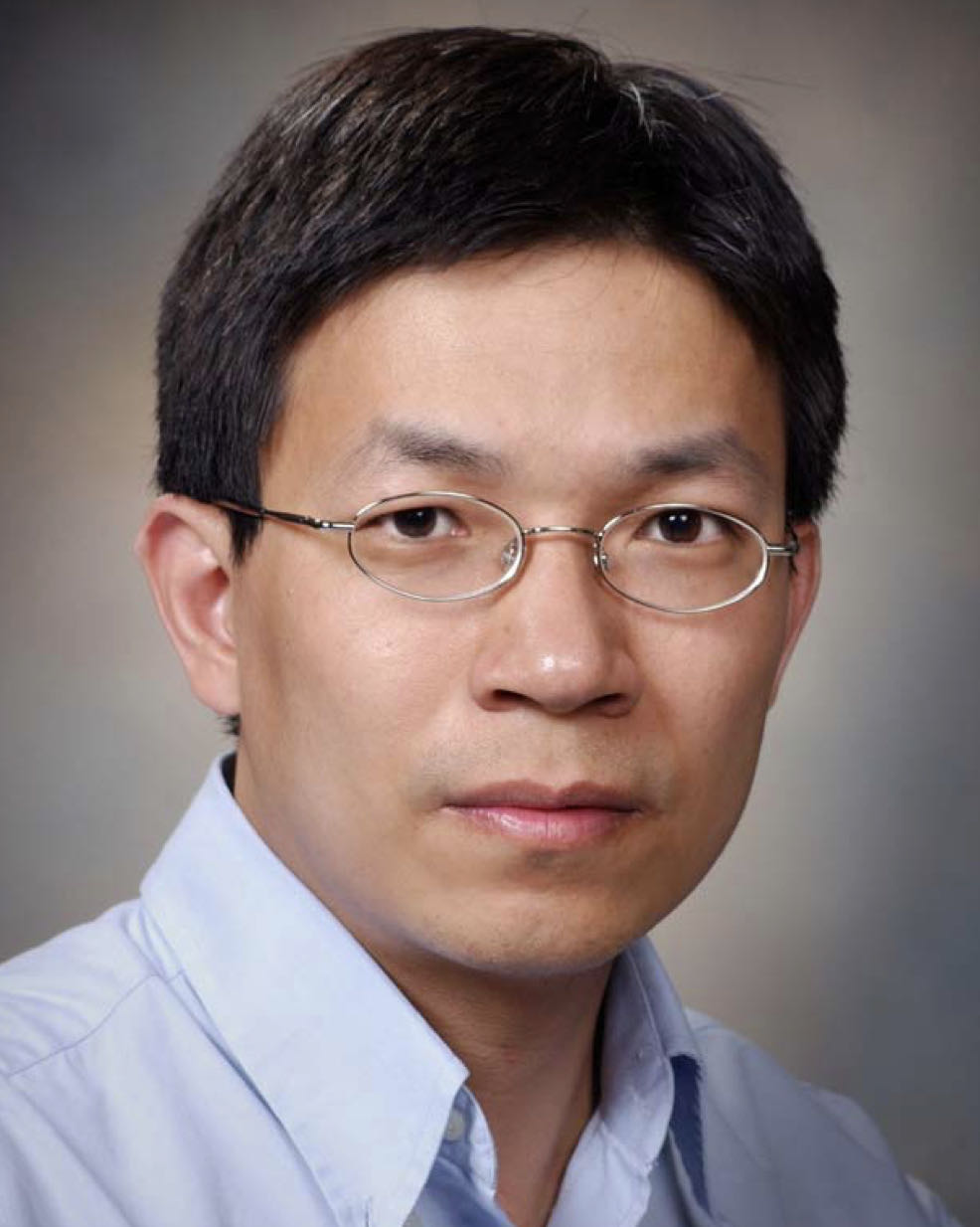}}]{Jie Liang} (Senior Member, IEEE)  received the B.E. and M.E. degrees from Xi'an Jiaotong University, China, the M.E. degree from National University of Singapore, and the PhD degree from the Johns Hopkins University, USA, in 1992, 1995, 1998, and 2003, respectively. From 2003 to 2004, he worked at the Video Codec Group of Microsoft Digital Media Division. Since May 2004, he has been with the School of Engineering Science, Simon Fraser University, Canada, where he is currently a Professor.
 
Jie Liang's research interests include Image and Video Processing, Computer Vision, and Deep Learning. He had served as an Associate Editor for several journals, including IEEE Transactions on Image Processing, IEEE Transactions on Circuits and Systems for Video Technology (TCSVT), and IEEE Signal Processing Letters. He has also served on three IEEE Technical Committees. He received the 2014 IEEE TCSVT Best Associate Editor Award, 2014 SFU Dean of Graduate Studies Award for Excellence in Leadership, and 2015 Canada NSERC Discovery Accelerator Supplements (DAS) Award.
\end{IEEEbiography}

\begin{IEEEbiography}[{\includegraphics[width=1in,height=1.25in,clip,keepaspectratio]{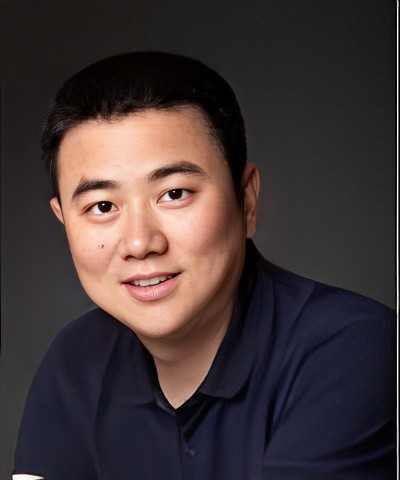}}]{Guohe Zhang} received the B.S. and Ph.D. degrees in electronics science and technology from Xi’an Jiaotong University, Shaanxi, China, in 2003 and 2008, respec- tively. He is currently an Associate Professor with the School of Microelectronics, Xi’an Jiaotong University. In 2009, he joined the School of Elec- tronic and Information Engineering, as a Lec- turer. He was promoted to an Associated Professor, in 2013. From 2009 to 2011, he had a three year’s
Postdoctoral Researcher with the School of Nuclear Science and Technology, Xi’an Jiaotong University. From February to May of 2013, he had a short term visiting to the University of Liverpool, U.K. His research interests fall in the area of semiconductor device physics and modeling, VLSI design and testing.
\end{IEEEbiography}

\begin{IEEEbiography}[{\includegraphics[width=1in,height=1.25in,clip,keepaspectratio]{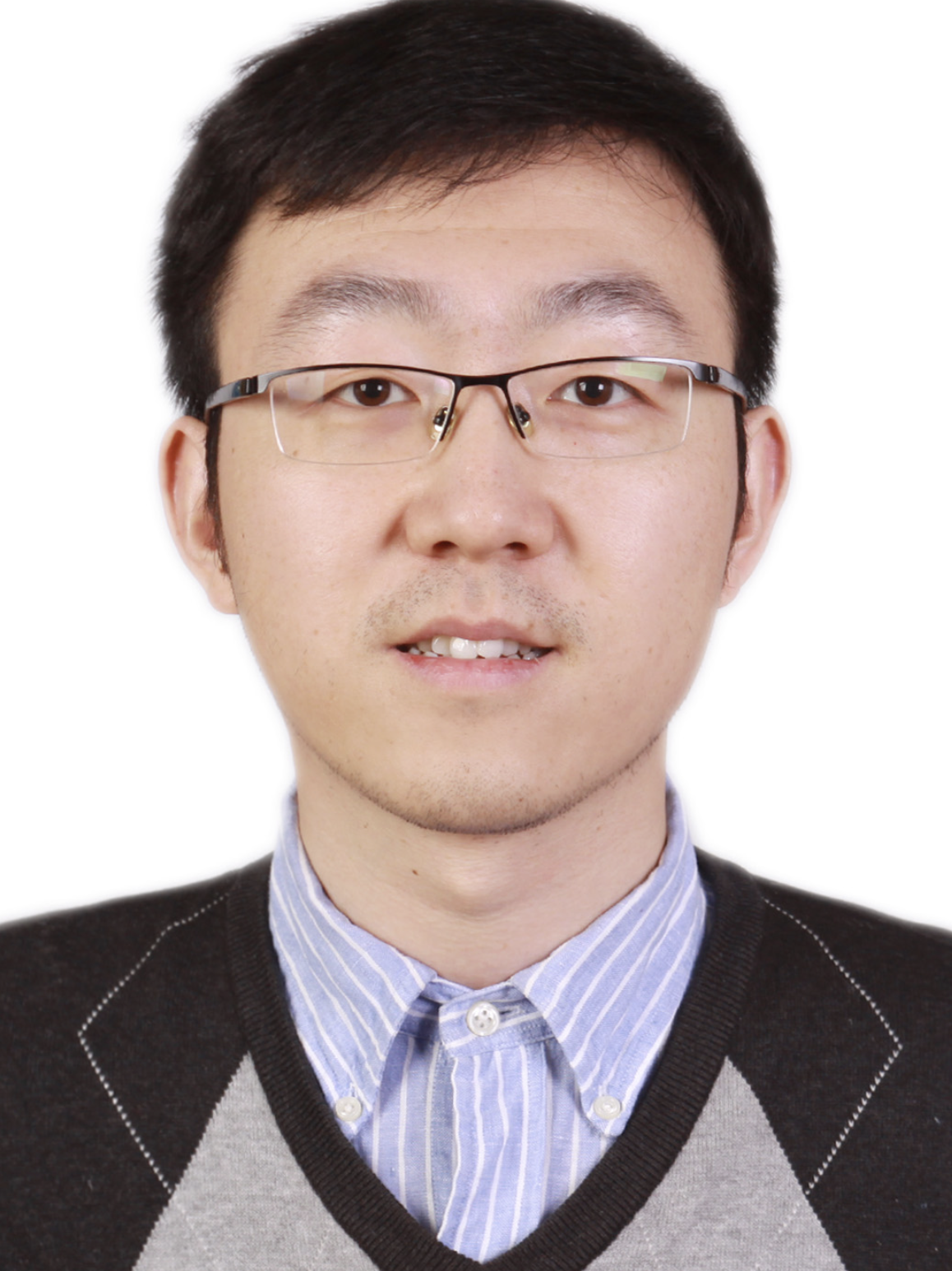}}]{Dong Liu } (Senior Member, IEEE) received the
BS and PhD degrees in electrical engineering from the University of Science and Technology of China (USTC), Hefei, China, in 2004 and 2009, respectively. He was a member of research staff with Nokia Research Center, Beijing, China, from 2009 to 2012. He joined USTC as an associate professor, in 2012. His research interests include
image and video coding, multimedia signal processing, and multimedia data mining. He has authored or coauthored more than 100 papers in international journals and conferences. He has 16 granted patents. He has one technical proposal adopted by AVS. He received the 2009 IEEE TRANSACTIONS ON CIRCUITS AND SYSTEMS FOR VIDEO TECHNOLOGY Best Paper Award and the VCIP 2016 Best 10 percent Paper Award. He and his team were winners of several technical challenges held in ICCV 2019, ACM MM 2018, ECCV 2018, CVPR 2018, and ICME 2016. He is a senior member of the CCF and CSIG, and an elected member of MSATC of IEEE CAS Society. He served as a registration co-chair for ICME 2019 and a symposium co-chair for WCSP 2014.
\end{IEEEbiography}

\begin{IEEEbiography}[{\includegraphics[width=1in,height=1.25in,clip,keepaspectratio]{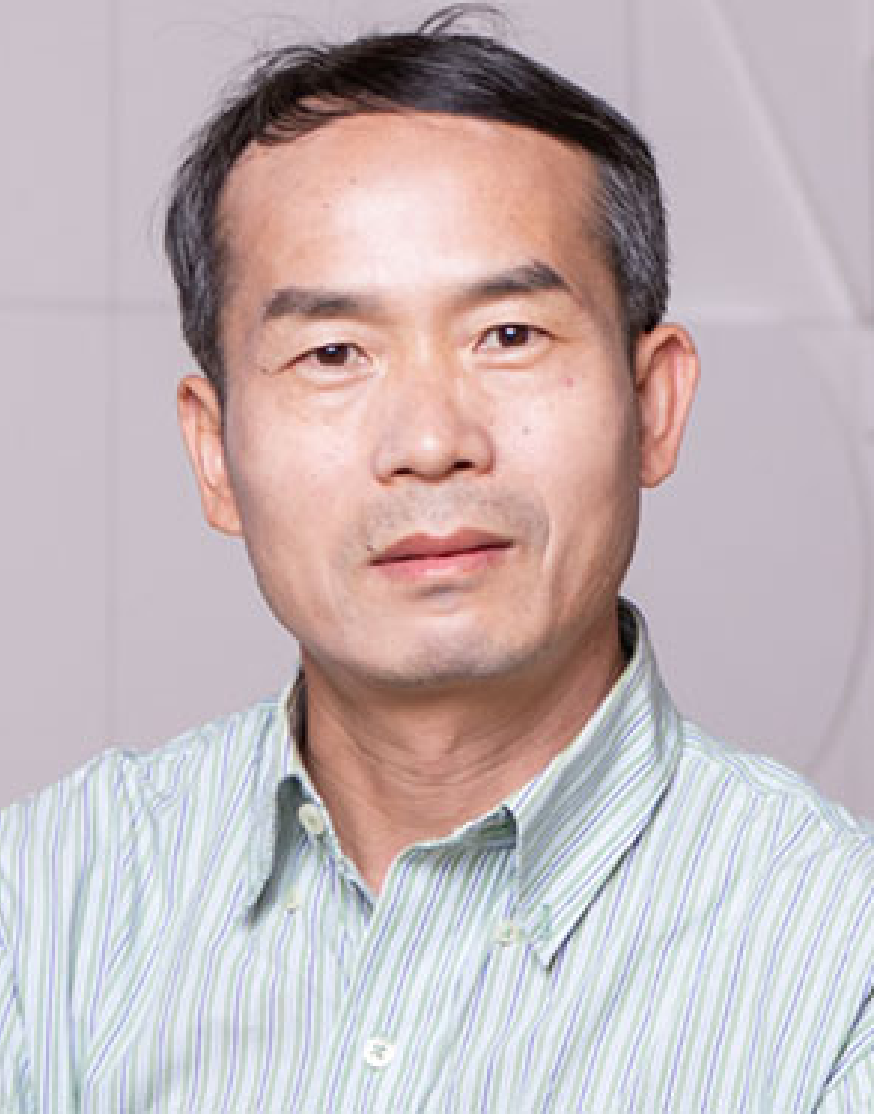}}]{Chengjie Tu} (Member, IEEE) received the B.S. degree in mechanical engineering from the University
of Science and Technology of China, Hefei, China,
in 1994, and the M.S. and Ph.D. degrees in electrical and computer engineering from Johns Hopkins
University, Baltimore, MD, USA, in 2001 and 2003,
respectively. He is currently the Chief Video Codec
Expert with Cloud Architecture and Platform Department, Tencent, Shenzhen, China. His research interests include image or video coding, processing, and
real time multimedia communication.
\end{IEEEbiography}

\begin{IEEEbiography}[{\includegraphics[width=1in,height=1.25in,clip,keepaspectratio]{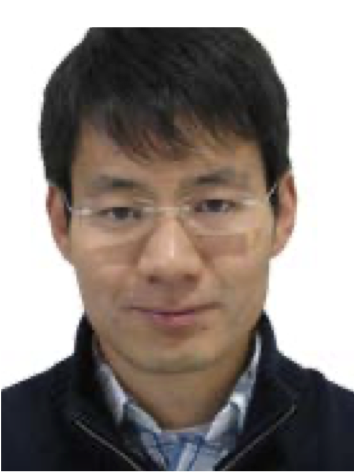}}]{Jingning Han} (Senior Member, IEEE) received the B.S. degree in electrical engineering from Tsinghua University, Beijing, China, in 2007, and the M.S. and Ph.D. degrees in electrical and computer engineering from the University of California at Santa Barbara, Santa Barbara, CA, USA, in 2008 and 2012, respectively.
He joined the WebM Codec Team, Google, Mountain View, CA, USA, in 2012, where he is the Main Architect of the VP9 and AV1 codecs, and leads the Software Video Codec Team. He has published more than 60 research articles. He holds more than 50 U.S. patents in the field of video coding. His research interests include video coding and computer science architecture. Dr. Han received the Dissertation Fellowship from the Department of Elec- trical and Engineering, University of California at Santa Barbara, in 2012. He was a recipient of the Best Student Paper Award at the IEEE International Conference on Multimedia and Expo, in 2012. He also received the IEEE Signal Processing Society Best Young Author Paper Award, in 2015.
\end{IEEEbiography}

\end{document}